\documentclass[11pt,a4paper]{article}
\pdfoutput=1
\usepackage{jheppub}
\newcommand{\rsm}{\ensuremath{{\rm SM}}}

\newcommand{\mSM}{\ensuremath{M_{H_{SM}}}}

\newcommand{\mw}{\ensuremath{M_{W}}}

\title{The Higgs boson in the minimal 3-3-1 model}

\author[a]{G. De Conto}%
\emailAdd{george.de.conto@gmail.com}
\affiliation[a]{
	Instituto  de F\'\i sica Te\'orica--Universidade Estadual Paulista \\
	R. Dr. Bento Teobaldo Ferraz 271, Barra Funda\\ S\~ao Paulo - SP, 01140-070,
	Brazil
}
\author[b]{A. C. B. Machado}%
\emailAdd{a.c.b.machado1@gmail.com}
\affiliation[b]{Laboratorio de F\'{\i}sica Te\'{o}rica e Computacional\\
	Universidade Cruzeiro do Sul -- Rua Galv\~{a}o Bueno 868 \\
	S\~{a}o Paulo, SP, 01506-000, Brazil}

\author[b]{J. P. B. C. de Melo}%

\abstract{In this paper we present the mass matrices and mass eigenstates for the CP-even neutral scalars in the minimal 331 model (m331) and its self-interactions, showing that the m331 automatically reproduces the Higgs potential of the Standard Model. We also present a method to generate numerical solutions for the quarks and leptons masses and their mixings, which we apply to study FCNC processes, being to calculate the contributions of all exotic neutral particles of the m331 to the mass differences in meson oscillations.}

\begin{document}

\maketitle

\section{Introduction}

In 2012 the Higgs boson was discovered at the Large Hadron Collider (LHC) on CERN \cite{higgs}, with its measured properties consistent with those of the Standard Model (SM) Higgs boson. Although a ten percent deviation in its couplings with the fermions is still possible, we can still count on the possibility of
the existence of additional scalar particles extending the scalar sector. Then, one question we can make is “how many doublet fields are there in the Higgs sector?”

We can consider, as examples of physics beyond the Standard Model, dark matter and/or the neutrino mass generation, and in many cases the extensions of the SM to explain these and other problems are made via new scalars. Therefore, after the SM-like Higgs boson discovery, the next challenge will be to measure its  mass and couplings to all known particles, including its couplings to $\rsm$
fermions and $\rsm$ gauge bosons, as well as the Higgs-particle self-couplings themselves. With these parameters measured it will be possible to distinguish if this is the $H_{\rsm}$ or something else. In this context the measurement of the Higgs boson self-couplings will be very relevant to establish the Higgs mechanism experimentally. It is well know that at the tree level, the self-couplings are uniquely determined by the
Higgs boson mass $\mSM$ and the vacuum expectation value of the Higgs boson
field $v_{SM}$, or equivalently the $W$ boson mass $\mw$ and the $SU(2)_L$ gauge
coupling $g$, since $v=2\mw/g$. More specifically, 
$\lambda_{HHH} = 3 \,M^2_{H_{SM}}/v$ and 
$\lambda_{HHHH} =3 \,M^2_{H_{SM}}/v^2$. So, if it is possible to adjust one of the scalars in the multi-higgs model to be the SM-higgs it is possible to obtain bonds that restrict the interactions of the extra scalars with the usual particles of the Standard Model, and thus the effect of flavor violation can be a way to detect new physics or to exclude models. In the present work, we will be concerned in making these adjustments to m331.

The minimal 3-3-1 model (m331) presents an extension to the scalar sector, with new charged and neutral scalars \cite{Pleitez}. Of those, only the neutral CP-even sector, with three neutral particles, has no analytical solution for its mass eigenstates and eigenvalues. So far, these particles had either been ignored or its mass matrix has been diagonalized by an orthogonal matrix dependent on three free parameters, where the phenomenological analyses had to take into account such unknown parameters. Also, the SM Higgs boson has to correspond to one of these CP-even scalars, usually assumed to be the lightest one of three. In our work we provide and analytical solution for this sector, where we can identify the Higgs boson and have only one unknown parameter that relates the symmetry and mass eigenstates of the other two CP-even scalars, with analytical expressions for all masses. Besides, we also show that the m331 Higgs has the same self-interactions as the SM Higgs, boosting our confidence that the m331 can replicate the SM and extend it. With this solution, the scalar sector of the m331 is now completely solved, with analytical solutions for all masses and eigenstates.

The addition of new particles brought to the SM by the m331 - not just of scalars, but also gauge bosons and fermions - brings into the Lagrangian other unknown parameters, more specifically, the matrices that diagonalize the fermion mass matrices. In the SM such matrices appear only in two instances, as matrix products, in the CKM and PMNS matrices, present in the interactions of quarks and leptons with the W boson, respectively. In the m331, these diagonalization matrices appear in the gauge-fermion interactions and the Yukawa sector as well, in combinations that do not correspond to the CKM and PMNS matrices. Therefore, these matrices have to be known in order to realize complete analyses involving such particles. In this work we also present a numerical method which allow us to find these matrices, in a manner that they generate the correct fermion masses and their products agree with the CKM and PMNS matrices.

With the solution for the CP-even sector and the method to find fermion diagonalization matrices we can now realize complete analyses within the m331. As an example of the use of these solutions we revisit a previous analysis \cite{machado1}. We calculate the contributions of all exotic neutral particles of the m331 to the mass differences in the meson oscillations $K^0 - \bar{K}^0$, $B^0 - \bar{B}^0$ and $D^0 - \bar{D}^0$. This time we are able to make a complete analysis of the problem, taking into account all particles of the model.

The outline of this paper is the following. In Sec.\ref{thesmhiggs} we briefly review the Higgs potential in the SM, in \ref{sec:model} we explain the features of the minimal 3-3-1 model, followed by Sec. \ref{sec:scalars2} we find the mass matrices and mass eigenstates for the neutral scalars. In Sec. \ref{the331higgs} we show that the m331 automatically reproduces the Higss potential of the SM. In Secs. \ref{sec:numericalsolutionsquarkmassmatrices} and \ref{sec:numericalsolutionleptonsmassmatrices}  we generate numerical solutions for the quarks and leptons masses and mixing.
Finally in Sec. \ref{massadiferences} we calculate the contributions of all exotic neutral particles of the m331 to the mass differences in the meson oscillations, followed by the conclusions.

\section{The Higgs boson in the Standard Model}
\label{thesmhiggs}

Here we briefly review the Higgs potential in the SM, since it is going to be relevant for the next sections, where we compare the CP-even scalar sector of the m331 with the SM scalar sector, in order to find an analytical solution to the former. In the SM, the Higgs doublet potential is given by
\begin{equation}
    V(\phi)=\mu^2 \phi^\dagger \phi +\frac{\lambda^2}{2} (\phi^\dagger \phi)^2
\end{equation}
where $\phi=(\phi^+ , \phi^0)^T$. The field $\phi^0$ develops a vacuum expectation value (VEV), and can be expressed as $\phi^0=(v_{SM}+H+Im(\phi^0))/\sqrt{2}$, where $v_{SM}$ is the VEV, $H$ is the Higgs boson and $Im(\phi^0)$ is the neutral Goldstone boson. Taking into account this expression, the potential then becomes
\begin{equation}\label{eq:PotHiggsSM}
    V(\phi)=\frac{m_H^2}{2} H H + \frac{m_H^2}{2 v_{SM}} H H H +\frac{m_H^2}{8 v_{SM}^2} H H H H + \cdots 
\end{equation}
where we used $\mu^2= - \lambda^2 v_{SM}^2/2$ and $\lambda = m_H/v_{SM}$. These identities come from the minimum condition of the potential vacuum and the resulting mass eigenstates. In the sections below we will show how the m331 relates to the potential described above.

\section{The minimal 3-3-1 model}
\label{sec:model}

Models with gauge symmetry $SU(3)_C\otimes SU(3)_L\otimes U(1)_X$ present new possibilities for the electroweak interactions. Here we consider the minimal 3-3-1 model (m331) in which there are new exotic quarks. Moreover, to give mass to all the particles, more scalar fields are needed. Hence, these models are intrinsically multi-Higgs models. 
In this model the electric charge operator is given by
$Q/|e|=T_{3}-\sqrt{3}T_{8}+X$,
where $e$ is the e electron charge, $T_{3,8}=\lambda_{3,8}/2$ (being $\lambda_{3,8}$ the Gell-Mann matrices) and $X$ is the hypercharge operator associated to the $U(1)_X$ group. In the sections below we will present the particle content of the model, for more details see refs. \cite{Pisano, Pleitez}. 

In the m331 the left-handed quark fields are chosen to form two antitriplets $Q^\prime_{mL}=(d_{m},\, -u_{m},j_{m})_{L}^{T}\sim({\bf 3}^{*},-1/3);\; m=1,2$; and a triplet $Q^\prime_{3L}=(u_{3},\, d_{3},\,
J)_{L}^{T} \sim({\bf 3},2/3)$. The right-handed ones are in singlets: $u_{\alpha
R}\sim({\bf 1},2/3)$, $d_{\alpha R}\sim({\bf
1},-1/3),\,\alpha=1,2,3$, $j_{mR}\sim({\bf 1},-4/3)$, and
$J_{R}\sim({\bf 1},5/3)$.
The scalar sector is composed by three triplets: $\eta=(\eta^0,\,\eta^{-}_1,\,\eta^+_2)^T\sim({\bf3},0)$,
$\rho=(\rho^+,\,\rho^0,\,\rho^{++})^T\sim({\bf3},1)$ and
$\chi=(\chi^-,\,\chi^{--},\,\chi^0)^T\sim({\bf3},-1)$, $\Psi_l=(\nu_l,\,l,\,l^c)^T_L\sim(\textbf{1},\textbf{3},0)$. Here we will add to the particle content of the m331 model, three sterile right-handed neutrinos, $\nu_{lR}\sim(\textbf{1},\textbf{1},0)$ ($l=e,\mu,\tau$).
The numbers in parenthesis are related to the transformation properties under $SU(3)_{L}$ and $U(1)_{X}$, respectively. For more details on interactions in the leptonic sector see Ref.\cite{Machado:2016jzb}.

The Yukawa interactions between fermions and scalars are given by:
\begin{eqnarray} 
-\mathcal{L}^q_Y &=& \bar{Q}_{mL} \left[ G_{m\alpha} U^{'}_{\alpha R} \rho^*+\tilde{G}_{m\alpha}D^{'}_{\alpha R} \eta^* \right]+
\bar{Q}_{3L} \left[ F_{3\alpha}U^{'}_{\alpha R} \eta + \tilde{F}_{3\alpha}D^{'}_{\alpha R} \rho \right] \nonumber \\&  +&
\bar{Q}_{mL}G'_{mi}j_{iR}\chi^* + \bar{Q}_{3L}g_J J_R \chi 
-\frac{1}{2}\epsilon_{ijk}\,\overline{(\Psi_{ia})^c}G^\eta_{ab} \Psi_{jb}\eta_k+\frac{1}{2\Lambda_S}\,
\overline{(\Psi_{a})^c} G^S_{ab} (\chi^*\rho^\dagger+ \rho^*\chi^\dagger)\Psi_{b} \nonumber \\ &+&
\overline{(\Psi_{aL})}G^\nu_{ab}\nu_{aR}\eta+ \overline{(\nu_{aR})^c}(M_R)_{ab}\nu_{bR}+H.c.,
\label{q1}
\end{eqnarray}
where $m=1,2$; $(a,b,\alpha)=1,2,3$; and $\Lambda_S$ is a mass scale generated by the effective interactions induced by the heavy scalar.
It means that FCNC processes in the lepton and quark sector are predictions of this model.

From Eq.(\ref{q1}) the quark mass matrices are given by
\begin{eqnarray}
M^u=&& G \frac{v_\rho}{\sqrt{2}}+F \frac{v_\eta}{\sqrt{2}} \\
M^d=&& \tilde{G} \frac{v_\eta}{\sqrt{2}}+\tilde{F}\frac{v_\rho}{\sqrt{2}}.
\end{eqnarray}

By choosing $v_\rho=54$ GeV and $v_\eta=240$ GeV the mixing between $Z$ and $Z^\prime$ vanishes independently of the value of $v_\chi$ (see the next section and Ref.~\cite{newp} for details). Also, these values for the vacuum expectation values guarantee, at tree level, the relation $M_Z=M_W/c_W$, where $c_W$ is the cosine of the Weinberg angle \cite{Dias:2006ns}.

The symmetry eigenstates $U'_{L,R}$, $D'_{L,R}$ (primed fields) and the mass eigenstates
$U_{L,R}$ $D_{L,R}$ (unprimed fields) are related by  $U'_{L,R}=(V^U_{L,R})^\dagger U_{L,R}$ and $D'_{L,R}=(V^D_{L,R})^\dagger D_{L,R}$, where $V^{U,D}_{L,R}$  are unitary matrices such that $V_L^{U}M^{u}V_R^{U\dagger}=\hat{M}^{u}$ and
$V_L^{D}M^{d}V_R^{D\dagger}=\hat{M}^{d}$, with $\hat{M}^{u}=diag(m_u,m_c,m_t)$ and $\hat{M}^{d}=diag(m_d,m_s,m_b)$. From those and Eq. \ref{q1} we also have $V_{CKM}=V_L^U V_L^{D\dagger}$. 

Finally, the most general potential, invariant under CP transformations, for the scalars is:
\begin{eqnarray}
\label{potencial_escalar}
V(\chi,\eta,\rho)&=& \sum_{i} \mu^2_{i} \phi^\dagger_i \phi_i
+\sum_{i=1,2,3}a_i(\phi^{\dagger}_i\phi_i)^2 +\sum_{m=4,5,6,i>j} a_m(\phi_i^\dagger \phi_i)(\phi_j^\dagger \phi_j)
\nonumber \\&+&
\sum_{n=7,8,9;i>j}a_n (\phi^{\dagger}_i\phi_j)(\phi^\dagger_j\phi_i) +( \alpha \,\epsilon_{ijk}\chi_{i}\rho_{j}\eta_{k}+h.c.) ,
\label{potential}
\end{eqnarray}
where we have used $\phi_1=\chi,\phi_2=\eta$ and $\phi_3=\rho$, except in the trilinear term. For more details of the scalar potential see Sec. \ref{sec:scalars2}.

\section{The scalar sector}
\label{sec:scalars2}

Taking the derivatives of Eq.~(\ref{potencial_escalar}) with respect to the vacua and setting these to zero we are able to
find expressions for $\mu^2_\chi$, $\mu^2_\eta$ and $\mu^2_\rho$. Using this we can find the mass matrices and mass eigenstates for all scalars. However, since in this work we are concerned solely with the CP-even neutral scalars, we will present the results only for this sector. The analytical results for the rest of the scalar sector can be found in \cite{DeConto:2014fza}.

\begin{equation}
    \mu_1^2 = -\frac{a_1 v_\chi^3+\frac{1}{2} a_4 v_\eta^2 v_\chi+\frac{1}{2} a_5 v_\rho^2 v_\chi-\frac{\alpha  v_\eta v_\rho}{\sqrt{2}}}{v_\chi}
\end{equation}

\begin{equation}
    \mu_2^2 = -\frac{a_2 v_\eta^3+\frac{1}{2} a_4 v_\eta v_\chi^2+\frac{1}{2} a_6 v_\eta v_\rho^2-\frac{\alpha  v_\rho v_\chi}{\sqrt{2}}}{v_\eta}
\end{equation}

\begin{equation}
    \mu_3^2 = -\frac{a_3 v_\rho^3+\frac{1}{2} a_5 v_\rho v_\chi^2+\frac{1}{2} a_6 v_\eta^2 v_\rho-\frac{\alpha  v_\eta v_\chi}{\sqrt{2}}}{v_\rho}
\end{equation}

\subsection{CP-even neutral scalars}\label{sec:CPeven}

The mass matrix for the CP-even scalars is given by

\begin{equation}
M_{h}=
\left(
\begin{array}{ccc}
2 a_2 v_\eta^2+\frac{v_\rho v_\chi \alpha }{\sqrt{2} v_\eta} & a_6 v_\eta v_\rho-\frac{v_\chi \alpha }{\sqrt{2}} & a_4 v_\eta v_\chi-\frac{v_\rho \alpha }{\sqrt{2}} \\
a_6 v_\eta v_\rho-\frac{v_\chi \alpha }{\sqrt{2}} & 2 a_3 v_\rho^2+\frac{v_\eta v_\chi \alpha }{\sqrt{2} v_\rho} & a_5 v_\rho v_\chi-\frac{v_\eta \alpha }{\sqrt{2}} \\
a_4 v_\eta v_\chi-\frac{v_\rho \alpha }{\sqrt{2}} & a_5 v_\rho v_\chi-\frac{v_\eta \alpha }{\sqrt{2}} & 2 a_1 v_\chi^2+\frac{v_\eta v_\rho \alpha }{\sqrt{2} v_\chi} \\
\end{array}
\right)
\end{equation}
in the basis $(X_{\eta}^0,X_{\rho}^0,X_{\chi}^0)^T M_h (X_{\eta}^0,X_{\rho}^0,X_{\chi}^0)$, where $X_{\phi}^0=Re(\phi^0)$. At the same time, we have the up quarks mass matrix (see Eq. \ref{q1}), coming from

\begin{eqnarray}
-\mathcal{L}^q_Y &&= \bar{U}'_L \left[ G \rho^0 + F \eta^0 \right]U'_R+\cdots\\&&=
\bar{U}'_L V^{U\dagger}_L V^U_L \frac{\left[ G(v_\rho+X_{\rho}^0+i I_{\rho}^0) + F(v_\eta+X_{\eta}^0+i I_{\eta}^0)\right]}{\sqrt{2}} V^{U\dagger}_R V^U_R U'_R+\cdots \\&&=
\bar{U}'_L V^{U\dagger}_L V^U_L \frac{\left[ Gv_\rho + Fv_\eta\right]}{\sqrt{2}} V^{U\dagger}_R V^U_R U'_R + \bar{U}_L V^{U}_L \frac{\left[ G X_{\rho}^0 + F X_{\eta}^0 \right]}{\sqrt{2}} V^{U\dagger}_R U_R \cdots \\&&=
\bar{U}_L \hat{M}^u U_R+ \bar{U}_L V^{U}_L \left[ G \frac{\lambda_{\rho^0}}{\sqrt{2}} + F \frac{\lambda_{\eta^0}}{\sqrt{2}} \right] V^{U\dagger}_R U_R H +\cdots
\end{eqnarray}
where $\hat{M}^u=V^{U\dagger}\left[ Gv_\rho + Fv_\eta\right]V^U_R=diag(m_u,m_c,m_t)$, $H$ is the SM higgs boson, and $\lambda_{\eta^0}$ and $\lambda_{\rho^0}$ are the projections of $X_{\eta}^0$ and $X_{\rho}^0$ onto $H$. If we want, at tree level, that the higgs couplings to the quarks in the m331 are the same as in the SM we need

\begin{equation}
V^{U}_L \left[ G \frac{\lambda_{\rho^0}}{\sqrt{2}} + F \frac{\lambda_{\eta^0}}{\sqrt{2}} \right]V^{U\dagger}_R= \frac{\hat{M}^u}{v_{SM}}=\frac{1}{v_{SM}}V^{U}_L \left[ G\frac{v_\rho}{\sqrt{2}} + F \frac{v_\eta}{\sqrt{2}}\right]V^{U\dagger}_R.
\end{equation}

Given that the matrices $G$ and $F$ are linearly independent (see Eqs. \ref{q1}, \ref{eq:matrizesG} and \ref{eq:matrizesF}), we have that $\lambda_{\eta^0}=v_\eta/v_{SM}$ and $\lambda_{\rho^0}=v_\rho/v_{SM}$, where $v_{SM}=\sqrt{v_\eta^2+v_\rho^2}$ (this identity for $v_{SM}$ can be seen from the expressions for the $Z$ and $W$ boson masses in the m331). This same line of reasoning can be applied to the Yukawa lagrangians for the down quarks and charged leptons, yielding the same results.

Only the $\rho^0$ and $\eta^0$ fields give mass to the known particles, therefore the SM higgs must be a linear combination of only these two fields, and since we know the projections we find that $H=\lambda_{\eta^0}X_\eta^0+\lambda_{\rho^0}X_\rho^0$. The CP-even mass matrix is symmetric and real, therefore it is diagonalized by orthogonal matrices as follows

\begin{equation}
R^T M_h R = diag(m_H^2,m_{h_1}^2,m_{h_2}^2)
\end{equation}
which imply
\begin{equation}
\left(\begin{array}{c}
H \\ h_1^0 \\ h_2^0
\end{array}\right)
=R^T\left(
\begin{array}{c}
X_\eta^0 \\ X_\rho^0 \\ X_\chi^0
\end{array}
\right)=
\left(
\begin{array}{ccc}
c_2 & c_1 s_2 & s_1 s_2 \\
-c_3 s_2 & c_1 c_2 c_3-s_1 s_3 & c_2 c_3 s_1+c_1 s_3 \\
s_2 s_3 & -c_3 s_1-c_1 c_2 s_3 & c_1 c_3-c_2 s_1 s_3 \\
\end{array}
\right)
\left(
\begin{array}{c}
X_\eta^0 \\ X_\rho^0 \\ X_\chi^0
\end{array}
\right)
\end{equation}
where $c_i$ and $s_i$ are the cosines and sines, respectively, of the angles $\theta_i$. To obtain $H=\lambda_{\eta^0}\eta^0+\lambda_{\rho^0}\rho^0$ we need that

\begin{equation}
c_2=\frac{v_\eta}{\sqrt{v_\eta^2+v_\rho^2}}, \qquad s_2=\frac{v_\rho}{\sqrt{v_\eta^2+v_\rho^2}}, \qquad \theta_1=0.
\end{equation}

The above identities leads us to

\begin{equation}
R=
\left(
\begin{array}{ccc}
\frac{v_\eta}{\sqrt{v_\eta^2+v_\rho^2}} & -\frac{v_\rho c_3}{\sqrt{v_\eta^2+v_\rho^2}} & \frac{v_\rho s_3}{\sqrt{v_\eta^2+v_\rho^2}} \\
\frac{v_\rho}{\sqrt{v_\eta^2+v_\rho^2}} & \frac{v_\eta c_3}{\sqrt{v_\eta^2+v_\rho^2}} & -\frac{v_\eta s_3}{\sqrt{v_\eta^2+v_\rho^2}} \\
0 & s_3 & c_3 \\
\end{array}
\right).
\label{eq:rotacaoescalaresCPpar}
\end{equation}
Now, we need the identity $R^T M_h R =  diag(m_H^2,m_{h_1}^2,m_{h_2}^2)$, which can be obtained if we set

\begin{eqnarray}
a_1=&&\nonumber-\frac{a_2 v_\eta^2 v_\rho^2}{v_\chi^2 \left(v_\eta^2-v_\rho^2\right)}+\frac{a_3 v_\eta^2 v_\rho^2}{v_\chi^2 \left(v_\eta^2-v_\rho^2\right)}-\frac{a_4 t_3 \left(v_\eta^3 \left(-ct^2_3\right)-v_\eta v_\rho^2 ct^2_3+v_\eta^3+v_\eta v_\rho^2\right)}{2 v_\rho v_\chi \sqrt{v_\eta^2+v_\rho^2}} \\&& -\frac{t_3}{4 v_\eta v_\rho v_\chi^3 \sqrt{v_\eta^2+v_\rho^2}} \Big[ \sqrt{2} \alpha  v_\eta^3 v_\rho v_\chi ct^2_3+\sqrt{2} \alpha  v_\eta^2 v_\rho^2 ct_3 \sqrt{v_\eta^2+v_\rho^2} \\&&\nonumber -\sqrt{2} \alpha  v_\eta^2 v_\chi^2 ct_3 \sqrt{v_\eta^2+v_\rho^2}-\sqrt{2} \alpha  v_\rho^2 v_\chi^2 ct_3 \sqrt{v_\eta^2+v_\rho^2}+\sqrt{2} \alpha  v_\eta v_\rho^3 v_\chi ct^2_3 \\&&\nonumber -\sqrt{2} \alpha  v_\eta^3 v_\rho v_\chi-\sqrt{2} \alpha  v_\eta v_\rho^3 v_\chi \Big]
\end{eqnarray}

\begin{equation}
a_5=\frac{\sqrt{2} \alpha  v_\eta}{v_\rho v_\chi}-\frac{a_4 v_\eta^2}{v_\rho^2},
\end{equation}

\begin{equation}
a_6= \frac{2 a_2 v_\eta^2}{v_\eta^2-v_\rho^2}+\frac{2 a_3 v_\rho^2}{v_\rho^2-v_\eta^2},
\end{equation}
where $t_3$ and $ct_3$ stands for tangent and cotangent of $\theta_3$, respectively. The above identities gives us the following expressions for the CP-even scalars masses:

\begin{equation}
m_H^2=\frac{2 a_2 v_\eta^4-2 a_3 v_\rho^4}{v_\eta^2-v_\rho^2},
\label{massaHiggs}
\end{equation}

\begin{equation}
m^2_{h_1}=\frac{-4 a_2 v_\eta^3 v_\rho^3+4 a_3 v_\eta^3 v_\rho^3-v_\eta t_3 (v_\eta-v_\rho) (v_\eta+v_\rho) \sqrt{v_\eta^2+v_\rho^2} \left(2 a_4 v_\eta v_\chi-\sqrt{2} \alpha  v_\rho\right)+\sqrt{2} \alpha  v_\chi \left(v_\eta^4-v_\rho^4\right)}{2 v_\eta v_\rho (v_\eta-v_\rho) (v_\eta+v_\rho)},
\end{equation}

\begin{equation}
m^2_{h_2}=\frac{-4 a_2 v_\eta^3 v_\rho^3+4 a_3 v_\eta^3 v_\rho^3+v_\eta ct_3 (v_\eta-v_\rho) (v_\eta+v_\rho) \sqrt{v_\eta^2+v_\rho^2} \left(2 a_4 v_\eta v_\chi-\sqrt{2} \alpha  v_\rho\right)+\sqrt{2} \alpha  v_\chi \left(v_\eta^4-v_\rho^4\right)}{2 v_\eta v_\rho (v_\eta-v_\rho) (v_\eta+v_\rho)}.
\end{equation}
Even though we have values for $v_\rho$, $v_\eta$ and $m_H$, we are not able to fix the other parameters in Eq. \ref{massaHiggs}. Therefore, the values for the other scalar masses are still undefined, since they depend on these parameters.

\section{The Higgs potential in the m331}
\label{the331higgs}

Taking the expressions for $\mu_i$, $v_{SM}$, $a_i$, $m_H$ and the projections of the $X_\eta^0$ and $X_\rho^0$ fields from the previous section, and substituting those in Eq. \ref{potencial_escalar}, we obtain
\begin{equation}\label{eq:PotHiggsm331}
    V(\chi,\eta,\rho)=\frac{a_2 v_\eta^4-a_3 v_\rho^4}{v_\eta^2-v_\rho^2} HH + \frac{a_2 v_\eta^4-a_3 v_\rho^4}{\left(v_\eta^2-v_\rho^2\right) \sqrt{v_\eta^2+v_\rho^2}} HHH + \frac{a_2 v_\eta^4-a_3 v_\rho^4}{4 v_\eta^4-4 v_\rho^4} HHHH \cdots
\end{equation}
Now, if we look at Eq. \ref{eq:PotHiggsSM}, term by term, the coefficients of the Higgs potential are
\begin{equation}
    \frac{m_H^2}{2} HH \rightarrow \frac{a_2 v_\eta^4-a_3 v_\rho^4}{v_\eta^2-v_\rho^2} HH
\end{equation}
\begin{equation}
    \frac{m_H^2}{2 v_{SM}} HHH \rightarrow \frac{a_2 v_\eta^4-a_3 v_\rho^4}{\left(v_\eta^2-v_\rho^2\right) \sqrt{v_\eta^2+v_\rho^2}} HHH
\end{equation}
\begin{equation}
    \frac{m_H^2}{8 v_{SM}^2} HHHH \rightarrow \frac{a_2 v_\eta^4-a_3 v_\rho^4}{4 v_\eta^4-4 v_\rho^4} HHHH.
\end{equation}
Comparing the above equations with Eqs. \ref{eq:PotHiggsm331} and \ref{massaHiggs}, and remembering that $v_{SM}=\sqrt{v_\eta^2+v_\rho^2}$, we see that the coefficients for each Higgs term match. Therefore, the m331 automatically reproduces the Higss potential of the SM if we impose that, at tree level, the Yukawa couplings for the Higgs boson in the m331 are the same as the ones in the SM.

\section{Generation of numerical solutions for the quark mass matrices and diagonalizations}\label{sec:numericalsolutionsquarkmassmatrices}

The new interactions in the m331, brought by its new particles, allow the diagonalization matrices of the fermionic sector to be present in the Lagrangian, in forms that are distinct from the CKM and PMNS matrices of the SM. Therefore, such diagonalization matrices have to be known so we can realize complete analyses within the m331. For the quark sector, the equations we have to solve, simultaneously, are the quarks mass matrices with their diagonalizations and the equation for the CKM matrix, as described below:

\begin{equation}
V_L^U M^u V_R^{U \dagger} = diag(m_u,m_c,m_t),
\label{eq:diagonalizacaoU}
\end{equation}

\begin{equation}
V_L^D M^d V_R^{D \dagger} = diag(m_d,m_s,m_b),
\label{eq:diagonalizacaoD}
\end{equation}

\begin{equation}
V_{CKM}=V_L^U V_L^{D \dagger},
\label{eq:CKM}
\end{equation}

where

\begin{equation}
M^u= G \frac{v_\rho}{\sqrt{2}}+F \frac{v_\eta}{\sqrt{2}},
\label{eq:massaU}
\end{equation}
\begin{equation}
M^d= \tilde{G} \frac{v_\eta}{\sqrt{2}}+\tilde{F}\frac{v_\rho}{\sqrt{2}},
\label{eq:massaD}
\end{equation}

\begin{equation}
G=\left(\begin{array}{ccc}
G_{11}&G_{12}&G_{13}\\
G_{21}&G_{22}&G_{23}\\
0 &0 &0 \\
\end{array}\right),\;\;\tilde{G}=
\left(\begin{array}{ccc}
\tilde{G}_{11}&\tilde{G}_{12} &\tilde{G}_{13} \\
\tilde{G}_{21} &\tilde{G}_{22}&\tilde{G}_{23}\\
0  &0 &0  \\
\end{array}\right),
\label{eq:matrizesG}
\end{equation}

\begin{equation}
F=\left(\begin{array}{ccc}
0&0&0\\
0&0&0\\
F_{31} &F_{32} &F_{33} \\
\end{array}\right),\;\;\tilde{F}=
\left(\begin{array}{ccc}
0&0 &0\\
0 &0 &0 \\
\tilde{F}_{31}  &\tilde{F}_{32}  &\tilde{F}_{33}  \\
\end{array}\right).
\label{eq:matrizesF}
\end{equation}

Our first step is to randomly generate unitary matrices for $V^U_L$, $V^U_R$ and $V^D_R$. To do so we use the following definition for a $3\times3$ unitary matrix \cite{PDita2001}
\begin{equation}
U=d_3(\phi_1,\phi_2,\phi_3) J_2(\theta_1) J_1(\theta_2) d_2(\phi_4,\phi_5) J_2(\theta_3) d_1(\phi_6)
\label{eq:definicaounitaria}
\end{equation}
where,
\begin{equation}
J_1(\theta)=
\left(
\begin{array}{ccc}
\cos (\theta) & -\sin (\theta) & 0 \\
\sin (\theta) & \cos (\theta) & 0 \\
0 & 0 & 1 \\
\end{array}
\right),
\end{equation}
\begin{equation}
J_2(\theta)=
\left(
\begin{array}{ccc}
1 & 0 & 0 \\
0 & \cos (\theta) & -\sin (\theta) \\
0 & \sin (\theta) & \cos (\theta) \\
\end{array}
\right),
\end{equation}
\begin{equation}
d_1(a)=diag\left(1,1,exp(i a)\right),
\end{equation}
\begin{equation}
d_2(a,b)=diag\left(1,exp(i a),exp(i b)\right),
\end{equation}
\begin{equation}
d_3(a,b,c)=diag\left(exp(i a),exp(i b),exp(i c)\right).
\end{equation}
Where we pick random values for the phases $\phi$ and the angles $\theta$ from 0 up to 2$\pi$. With that we have $V^U_L$, $V^U_R$ and $V^D_R$ numerically fixed, then we can now find $V^D_L$ from Eq. \ref{eq:CKM} as
\begin{equation}
V_L^D=V_{CKM}^\dagger V_L^U,
\end{equation}
and with that we now have all the diagonalization matrices. With those in hand we find, using Eqs. \ref{eq:diagonalizacaoU} and \ref{eq:diagonalizacaoD}, the mass matrices to be
\begin{equation}
M^u=V_L^{U\dagger} diag(m_u,m_c,m_t) V_R^U\qquad, \qquad M^d=V_L^{D\dagger} diag(m_d,m_s,m_b) V_R^D.
\end{equation}
Finally, given that we have numerical solutions for $M^u$ and $M^d$, and we are working with $v_\eta=$ 240 GeV and $v_\rho=$ 54 GeV, we use Eqs. \ref{eq:massaU} and \ref{eq:massaD} to find the matrices $G$, $F$, $\tilde{G}$ and $\tilde{F}$ by direct inspection. It can be easily done because $G$ and $F$ are linearly independent, with the same happening for the matrices $\tilde{G}$ and $\tilde{F}$.

To summarize the process:
\begin{itemize}
	\item Generate randomly the matrices $V^U_L$, $V^U_R$ and $V^D_R$ using Eq. \ref{eq:definicaounitaria},
	\item Find $V_L^D$ using $V_L^D=V_{CKM}^\dagger V_L^U$,
	\item Find $M^u$ and $M^d$ using $M^u=V_L^{U\dagger} diag(m_u,m_c,m_t) V_R^U$ and $M^d=V_L^{D\dagger} diag(m_d,m_s,m_b) V_R^D$,
	\item Find $G$, $F$, $\tilde{G}$ and $\tilde{F}$ by direct inspection using Eqs. \ref{eq:massaU} and \ref{eq:massaD}.
\end{itemize}
Repeating the procedure described here we are able to generate as many numerical solutions as necessary.

\section{Generation of numerical solutions for the lepton mass matrices and diagonalizations}
\label{sec:numericalsolutionleptonsmassmatrices}

The situation for the leptons is similar to the quark case. Here the equations to be simultaneously solved are

\begin{equation}
V_L^{l\dagger} M^l V_R^L=diag(m_e,m_\mu,m_\tau)
\label{eq:diagonalizacaoleptons}
\end{equation}
\begin{equation}
V_L^{\nu \dagger} M^\nu V_L^\nu = diag(m_{\nu 1},m_{\nu 2},m_{\nu 3})
\label{eq:diagonalizacaoneutrinos}
\end{equation}
\begin{equation}
V_{PMNS}=V_L^{l \dagger} V_L^\nu
\label{eq:pmns}
\end{equation}
where
\begin{equation}
M^l=G^\eta \frac{v_\eta}{\sqrt{2}}+\frac{1}{\Lambda_S}G^S v_\rho v_\chi=G^\eta \frac{v_\eta}{\sqrt{2}}+G^S v_\rho
\label{eq:matrizmassaleptons}
\end{equation}
\begin{equation}
M^\nu=-\frac{v_\eta^2}{2} G^\nu \frac{\bar{M}}{M_R} G^{\nu \dagger}=-\bar{G}^{\nu}\bar{M} \bar{G}^{\nu  \dagger}, \qquad \bar{M}=diag(\bar{r}_1,\bar{r}_2,1), \qquad \bar{r}_i=\frac{M_R}{M_i}.
\label{eq:matrizmassaneutrinos}
\end{equation}
Notice that in Eq. \ref{eq:diagonalizacaoneutrinos} both diagonalization matrices are left, while in Eq. \ref{eq:diagonalizacaoleptons} we have two different matrices, left and right. In Eq. \ref{eq:matrizmassaleptons} we assume an effective operator for the lepton masses dependent on a parameter $\Lambda_S\sim v_\chi$, which leads us to the right side of same equation. 

For the leptons, the matrices $V_{L,R}^l$ and $V_L^\nu$ are unitary. Also, $G^\eta$ is an antisymmetric matrix, $G^S$ a symmetric one and $G^\nu$ is an arbitrary complex matrix. To find a numerical solution for those, we first generate randomly the matrices $V^\nu_L$ and $V_R^l$ using the definition from Eq. \ref{eq:definicaounitaria}, picking random numbers between 0 and 2$\pi$ for the phases and angles. Then, from eq. \ref{eq:pmns}, we find $V_L^l=(V_{PMNS} V_L^{\nu \dagger})^\dagger$, which leaves us with all diagonalization matrices at hand.

From Eq. \ref{eq:matrizmassaleptons} we find $M^l=V_L^l diag(m_e,m_\mu,m_\tau)V_R^{l \dagger}$. We also have that $G^\eta$ is antisymmetric and $G^S$ is symmetric and $M^l$ is given by a linear combination of those (see Eq. \ref{eq:matrizmassaleptons}), therefore we can obtain them by decomposing $M^l$ in its symmetric and antisymmetric components:
\begin{equation}
G^\eta \frac{v_\eta}{\sqrt{2}}=\frac{1}{2} \left( M^l - M^{lT} \right),
\end{equation}
\begin{equation}
G^S v_\rho = \frac{1}{2} \left( M^l + M^{lT} \right).
\end{equation}
Which gives us a complete solution for the charged leptons mass matrices and diagonalizations.

For the neutrinos we randomly generate the $\bar{M}$ matrix, with $\bar{r}_1$ and $\bar{r}_2$ real and in the $[0,1]$ interval. Then we numerically find the solution for the $\bar{G}^{\nu }$ matrix using Eq. \ref{eq:matrizmassaneutrinos}, giving us solutions to the entire leptonic sector, where $M_R$ remains a free parameter to be adjusted.

\section{Mass differences of the $K^0 - \bar{K}^0$, $B_s^0 - \bar{B}_s^0$ and $D^0_d - \bar{D}^0_d$ systems in the minimal 3-3-1 model}
\label{massadiferences}

Given the results presented in the previous sections, we now apply them to calculate the mass differences of the $K^0 - \bar{K}^0$, $B_s^0 - \bar{B}_s^0$ and $D^0_d - \bar{D}^0_d$ systems. This subject has already been explored in a previous work \cite{machado1}, however, at the time the solution for the CP-even scalar sector was unavailable. This made the previous work partially complete, because not all contributions from the model were studied. Now that we have such solution, and that we may generate as many diagonalization matrices as needed (unlike the previous work), we can make a complete analysis of these mass differences in the minimal 3-3-1 model.

\subsection{Theoretical calculations for $\Delta m_K$, $\Delta m_{B}$ and $\Delta m_D$}\label{sec:theoreticalcalculations}

The FCNC processes that we will consider here are induced by the $Z^\prime$, neutral scalars ($h_1$ and $h_2$) and pseudoscalars ($A^0$) present in the m331. The neutral currents mediated by the $Z^\prime$ have the following interactions to quarks: 
\begin{eqnarray}
\mathcal{L}_{Z^\prime}&=&-\frac{g}{2\cos\theta_W}\sum_{q=U,D}[\bar{q}_L\gamma^\mu K^q_L q_L+
\bar{q}_R\gamma^\mu K^q_R q_R]Z^\prime_\mu,
\label{zprime}
\end{eqnarray}
where we have defined
\begin{equation}
K^q_L=V^{q}_LY^q_LV^{q\dagger}_L,\;\; K^q_R =V^{q}_RY^q_RV^{q\dagger}_R,\;\;q=U,D;
\label{kas}
\end{equation}
and
\begin{equation}
Y^U_L=Y^D_L=-\frac{1}{2\sqrt{3}h(x)}\textrm{diag}\,[ -2(1-2x),-2(1-2x),1]
\label{yudl}
\end{equation}
and
\begin{equation}
Y^U_R=-\frac{4x}{\sqrt{3}h(x)}\,\textbf{1}_{3\times3} ,\quad
Y^D_R=\frac{2x}{\sqrt{3}h(x)}\textbf{1}_{3\times3}
\label{yudr}
\end{equation}
where $h(x)\equiv(1-4x)^{1/2},\;x=\sin^2\theta_W$. See Ref.~\cite{dpp}.

Note that, since $Y^{U,D}_R$ \textbf{is} proportional to the identity matrix, there are no FCNCs in the right-handed currents coupled to the $Z^\prime$.

The neutral currents mediated by scalars and pseudoscalars are obtained from the Yukawa interactions in Eq.(\ref{q1}), and are as follows:
\begin{equation}
-\mathcal{L}_{qqh}=\sum_{q=U,D}\overline{q}_L\mathcal{K}^q q_R
+\textrm{mass terms}+H.c. ,
\label{yukan}
\end{equation}
where we have defined $\mathcal{K}^U= V^U_L\mathcal{Z}^UV^{U\dagger}_R$ and $\mathcal{K}^D= V^D_L\mathcal{Z}^D
V^{D\dagger}_R$. These interactions in matrix form are (in the quark mass eigenstates  basis):
\begin{eqnarray}
\mathcal{Z}^U=\left(\begin{array}{ccc}
G_{11}\rho^0 &G_{12}\rho^0&G_{13}\rho^0\\
G_{21}\rho^0&G_{22}\rho^0&G_{23}\rho^0\\
F_{31}\eta^0&F_{32}\eta^0&F_{33}\eta^0
\end{array}
\right),\;\;
\mathcal{Z}^D=\left(\begin{array}{ccc}
\tilde{G}_{11}\eta^0&\tilde{G}_{12}\eta^0&\tilde{G}_{13}\eta^0\\
\tilde{G}_{21}\eta^0&\tilde{G}_{22}\eta^0&\tilde{G}_{23}\eta^0\\
\tilde{F}_{31}\rho^0&\tilde{F}_{32}\rho^0&\tilde{F}_{33}\rho^0
\end{array}
\right),
\label{yukanma}
\end{eqnarray}
where $\eta^0$ and $\rho^0$ are still symmetry eigenstates. These neutral symmetry
eigenstates  may be written as $\sqrt{2}x^0=\textrm{Re}
\,x^0+i\textrm{Im}\,x^0$, with $x^0=\eta^0,\rho^0$, and their relations to mass eigenstates are shown in the appendix \ref{sec:CPodd} for the CP-odd scalars and section \ref{sec:CPeven} for the CP-even scalars.

\subsection{$\Delta F=2$ processes}
\label{sec:deltaf2}

\subsubsection{$\Delta M_K$}
\label{subsec:mk}

As it is already well known, the $\Delta M_K$ mass difference in the SM is given by
$\Delta M^{SM}_K=\zeta^{SM}_{ds}\langle \bar{K}^0\vert (\bar{s}d)^2_{V-A} \vert K^0\rangle$ where,
using only the $c$-quark contribution, we have
\begin{equation}
\zeta^{SM}_{ds}=\frac{1}{3}M_Kf^2_K \frac{G^2_Fm^2_c}{16\pi^2}\,[(V_{CKM})^*_{us}(V_{CKM})_{ud}]^2=(2.83\pm0.11 )\times 10^{-16}\,\textrm{GeV},
\label{deltaksm}
\end{equation}
where we have neglected QCD corrections, and used the vacuum insertion approximation with $f_K=0.1598$ GeV \cite{branco}. For the other experimental values we used \cite{PDG}.

Now considering the contribution of the extra neutral vector boson.
From Eq.~(\ref{zprime}), the effective $Z^\prime$ interaction Hamiltonian inducing the $K^0 \to \bar{K}^0$ transition,
at the tree level, is given by
\begin{eqnarray}
\mathcal{H}^{\Delta S=2}_{eff}\vert_{Z^\prime}=\frac{g^2}{ 4 c^2_W M^2_{Z^\prime}} [\bar{s}_L (K^D_L)_{sd}
\gamma^\mu d_L]^2,
\label{heff}
\end{eqnarray}
which gives us the following extra contribution:
\begin{eqnarray}
\Delta M_K\vert_{Z^\prime}=2\textrm{Re}\,\langle \bar{K}^0\vert \mathcal{H}^{\Delta S=2}_{eff}
\vert_{Z^\prime}\vert K^0\rangle= \textrm{Re}\,\zeta^{Z^\prime}_{ds}\,\langle \bar{K}^0\vert (\bar{s}d)^2_{V-A}
\vert K^0\rangle
\label{deltak}
\end{eqnarray}
where
\begin{equation}
\textrm{Re}\,\zeta^{Z^\prime}_{ds}=\textrm{Re}\,\frac{G_F}{2 \sqrt{2} c^2_W}\,\frac{M^2_W}{M^2_{Z^\prime}}
\;[(K^D_L)_{ds}]^2.
\label{deltak331}
\end{equation}

The CP-even scalar contributions come from Eq.~(\ref{yukan}), and we considered the interactions between the  $d$ and $s$ quarks mediated by $\textrm{Re}x^0_i$,  giving
\begin{eqnarray}
-\mathcal{L}_{dsh}&=&
\frac{1}{\sqrt2}\sum_i[(I^{i}_K)_{ds} \bar{s_L}d_R \textrm{Re} x^0_i+ (I^{i*}_K)_{sd}\bar{d}_Ls_R \textrm{Re} x^0_i + H.c.]
\nonumber\\&=&\frac{1}{2 \sqrt2} \sum_i[(I^{i+}_K)_{ds}(\bar{d}s)+(I^{i-}_K)_{ds}(\bar{d}\gamma_5s)]\textrm{Re} x^0_i+ H.c.,
\label{newint}
\end{eqnarray}
where, $(I^i_K)_{q_1q_2} = (\mathcal{K}^D)_{q_1q_2} R_{x i}$ - being $R_{x i}$ the matrix that relates the CP-even scalar symmetry and mass eigenstates (see Eq. \ref{eq:rotacaoescalaresCPpar}) - with $x=\eta,\rho$ and $q_1,q_2=d,s$ for the CP-even scalars and quarks, respectively;
$i$ runs over the CP-even neutral scalar mass eigenstates. We also define  $(I^{i\pm}_K)_{ds} =(I^{i}_K)_{ds} \pm (I^{i*}_K)_{sd}$. 

The interaction for $CP$-odd fields are the given in the same Lagrangian of the $CP$-even, see Eq. (\ref{newint}), but with $\textrm{Re}x^0_i\to \textrm{Im}x^0_i$ and $(I^i_K)_{q_1q_2} \to(I^i_K)^A_{q_1q_2}
= (\mathcal{K}^D)_{q_1q_2} V_{x i}$, where $V_{x i}$ relates the CP-odd scalar symmetry and mass eigenstates (see Eq. \ref{a4}).

The effective Hamiltonian induced by Eq.~(\ref{newint}) and the respective contribution of the pseudoscalar $\textrm{Im} x^0_1$
to the $K^0 \leftrightarrow \bar{K}^0$ transition is:
\begin{eqnarray}
\mathcal{H}^{\Delta S=2}_{eff}\vert_{h+A}&=&\sum_i\frac{1}{8 m^2_i}[(I^{i+}_K)^2_{ds}(\bar{s}d)^2+(I^{i-}_K)^2_{ds}(
\bar{s}\gamma_5d)^2]\nonumber \\&-&
\sum_i\frac{1}{8 m^2_A}[[(I^{i+}_K)^A_{ds}]^2(\bar{s}d)^2+[(I^{i-}_K)^A_{ds}]^2(\bar{s}\gamma_5d)^2.
\label{heff2}
\end{eqnarray}
Defining as usual
\begin{equation}
\Delta M_K\vert_{h,A}=2
\langle \bar{K}^0\vert
\mathcal{H}^{\Delta S=2}_{eff}\vert_{h,A} \vert K^0\rangle=\textrm{Re}\zeta^{h,A}_{sd}\langle \bar{K}^0
\vert (\bar{s}d)^2_{V-A} \vert K^0\rangle,
\label{effective2}
\end{equation}
and  using the matrix elements~\cite{branco}:
\begin{eqnarray}
&&\langle \bar{K}^0\vert (\bar{s}d)(\bar{s}d)\vert K^0\rangle=-\frac{1}{4}\left[1-\frac{M^2_K}{(m_s+m_d)^2}
\right] \langle
\bar{K}^0\vert (\bar{s}d)^2_{V-A} \vert K^0\rangle,\nonumber \\&&
\langle \bar{K}^0\vert (\bar{s}\gamma_5d)(\bar{s}\gamma_5d)\vert K^0\rangle=\frac{1}{4}\left[1-1
1\frac{M^2_K}{(m_s+m_d)^2}
\right]\langle \bar{K}^0\vert (\bar{s}d)^2_{V-A} \vert K^0\rangle,
\label{v1v2}
\end{eqnarray}
we find
\begin{eqnarray}
\textrm{Re}\zeta^h_{ds}\!\!=\!\!\textrm{Re}\sum_i\frac{1}{32 m^2_i}\left(-(I^{i+}_K)^2_{ds}\left[1\!\!-
\!\!\frac{M^2_K}{(m_s+m_d)^2} \right]\!\!
+\!\!(I^{i-}_K)^2_{ds}\left[1\!\!-\!\!\frac{11M^2_K}{(m_s+m_d)^2} \right] \right)\textrm{GeV}^{-2}.
\label{zetas2}
\end{eqnarray}

We have similar expressions for the pseudoscalar contributions by making
$I^{i\pm}_K\to (I^{i\pm})^A_K$, with
$R_{\eta1}\to V_{\eta1}$, $R_{\rho1}\to V_{\rho1}$ and $m_i\to m_{Ai}$.  Thus, the $\Delta M_K$ in the present
model includes $Z^\prime$
and neutral scalar and pseudoscalar contributions,
\begin{equation}
\Delta M_K\vert_{331}= 
\Delta M_K\vert^{Z^\prime}+\Delta M_K\vert^{h1}+\Delta M_K\vert^{h2}+\Delta M_K\vert^A\equiv\zeta_{331}\langle \bar{K}^0
\vert (\bar{s}d)^2_{V-A} \vert K^0\rangle,
\label{full}
\end{equation}
with $\zeta^{331}_{ds}=\zeta^{Z^\prime}_{ds}+\zeta^{h1}_{ds}+\zeta^{h2}_{ds}+\zeta^A_{ds}$, $\langle \bar{K}^0
\vert (\bar{s}d)^2_{V-A} \vert K^0\rangle=f_K^2 M_K$. Therefore, to satisfy the experimental results we need that $\zeta^{331}_{ds}+\zeta^{SM}_{ds}=\Delta M_K^{exp}$. We need a similar result for the other mass differences (which we will discuss next), where the sum of the respective contributions from the m331 and the SM gives us the experimental results.

\subsubsection{$\Delta M_{B}$}
\label{subsec:mb}

Considering the oscillation $B^0_s-\bar{B}^0_s$ mass difference we have $\Delta M^{SM}_B=(1.20 \pm 0.18) \times 10^{-11}$ GeV , where it was used $f_B=0.216\pm 0.015$ GeV \cite{Artuso:2015swg}. Similar to the previous case, to obtain the m331 contributions for this process we have to change the $d$ and $s$ quarks to $s$ and $b$, respectively, in the expressions for $\zeta^{Z^\prime}_{ds}$, $\zeta^{h1}_{ds}$, $\zeta^{h2}_{ds}$ and $\zeta^A_{ds}$.

\subsubsection{$\Delta M_D$}
\label{subsec:md}

Considering the oscillation $D^0_d-\bar{D}^0_d$ mass difference, the SM model prediction is given by
\begin{eqnarray}
\Delta M^{SM}_D &=& \frac{4 G^2_F}{3 \pi^2} \zeta_{s}\zeta_{d} \frac{(m_s^2 - m_d^2)^2}{m_c^2} f_D^2 m_D (B_D - 2 B^\prime_D) =(0.024 \pm 4.914)\times 10^{-16}\, \textrm{GeV}
 \end{eqnarray}
with $f_D = 0.165$ GeV, $B_D = B^\prime_D = 1$, and $\zeta_{i} = V_{ic}^* V_{iu}$. \cite{Datta:1984jx}. Similar to the $\Delta M_K$ case, to obtain the m331 contributions for this process we have to change the $d$ and $s$ quarks to $c$ and $u$, respectively, in the expressions for $\zeta^{Z^\prime}_{ds}$, $\zeta^{h1}_{ds}$, $\zeta^{h2}_{ds}$ and $\zeta^A_{ds}$.

\subsection{Individual contributions from the exotic particles}\label{sec:individualcontributions}

In this section we shall study the individual contributions from each exotic particle to the mesons' mass differences. When taking into account all contributions we are left with several parameters to explore. First, when considering the quark sector we are left with three of the diagonalization matrices free, given that the quark mass matrices and one of the diagonalization matrices can be written as functions of the CKM matrix and the three free diagonalization matrices (see section \ref{sec:numericalsolutionsquarkmassmatrices}). This leaves us with 27 free parameters, six phases and three angles coming from each of the three diagonalization matrices. Second, we have $\theta_3$, the angle that relates the symmetry and mass eigenstates for the CP-even scalars (section \ref{sec:scalars2}), and the vacuum expectation value $v_\chi$, which dictates the mass of the $Z'$ prime boson and the diagonalization matrix of the CP-odd sector; giving us two more free parameters. Finally, the masses of the scalar particles (two CP-even and one CP-odd), giving us another three parameters, leading to a total of 32 free parameters to explore. By varying $v_\chi$, which controls the $Z'$ mass and the CP-odd projections of the symmetry eigenstates over mass eigenstates, varying $\theta_3$, which controls the projections of the CP-even scalars, and by considering different solutions for the quark diagonalization matrices (shown in appendix \ref{sec:numericalsolutionsused}), we see the behavior of the $\zeta^a_{bc}$ functions presented in the previous section.

\subsubsection{$Z'$ gauge boson}

\begin{figure}
\centering
\includegraphics[width=\linewidth]{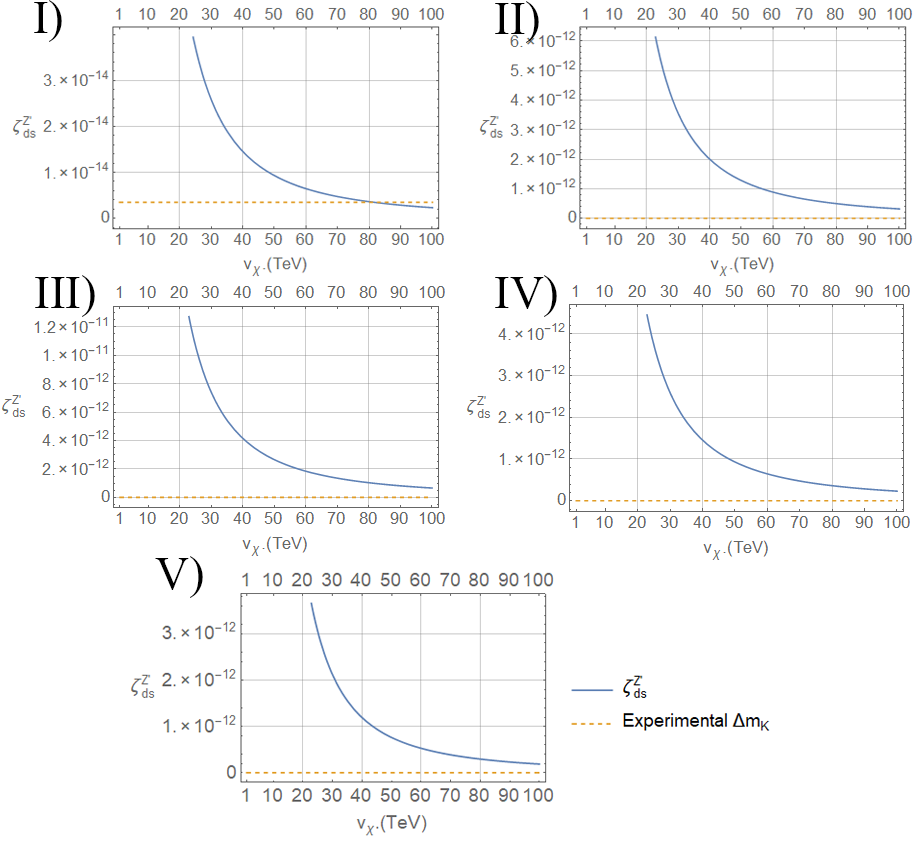}
\caption{Contributions to $\Delta m_K$ from the $Z'$ boson. The numbers I-V correspond to the numerical solutions presented in appendix \ref{sec:numericalsolutionsused}.}
\label{fig:ZetaZprime_ds_Todos}
\end{figure}

\begin{figure}
	\centering
	\includegraphics[width=\linewidth]{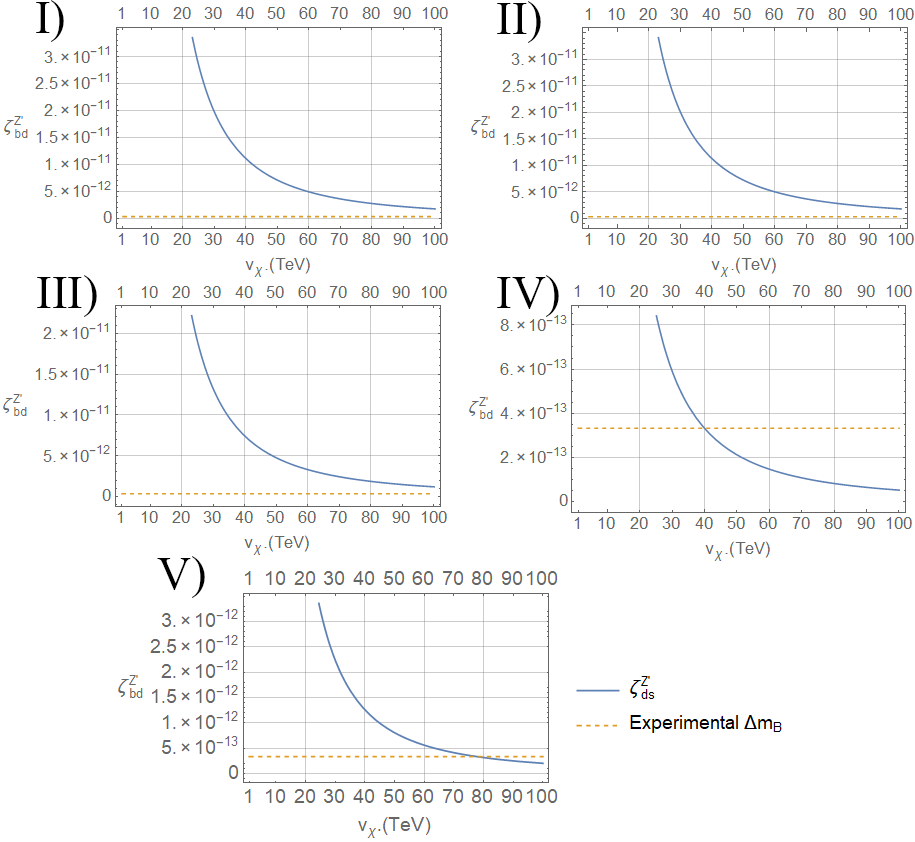}
	\caption{Contributions to $\Delta m_B$ from the $Z'$ boson. The numbers I-V correspond to the numerical solutions presented in appendix \ref{sec:numericalsolutionsused}.}
	\label{fig:ZetaZprime_bd_Todos}
\end{figure}

\begin{figure}
	\centering
	\includegraphics[width=\linewidth]{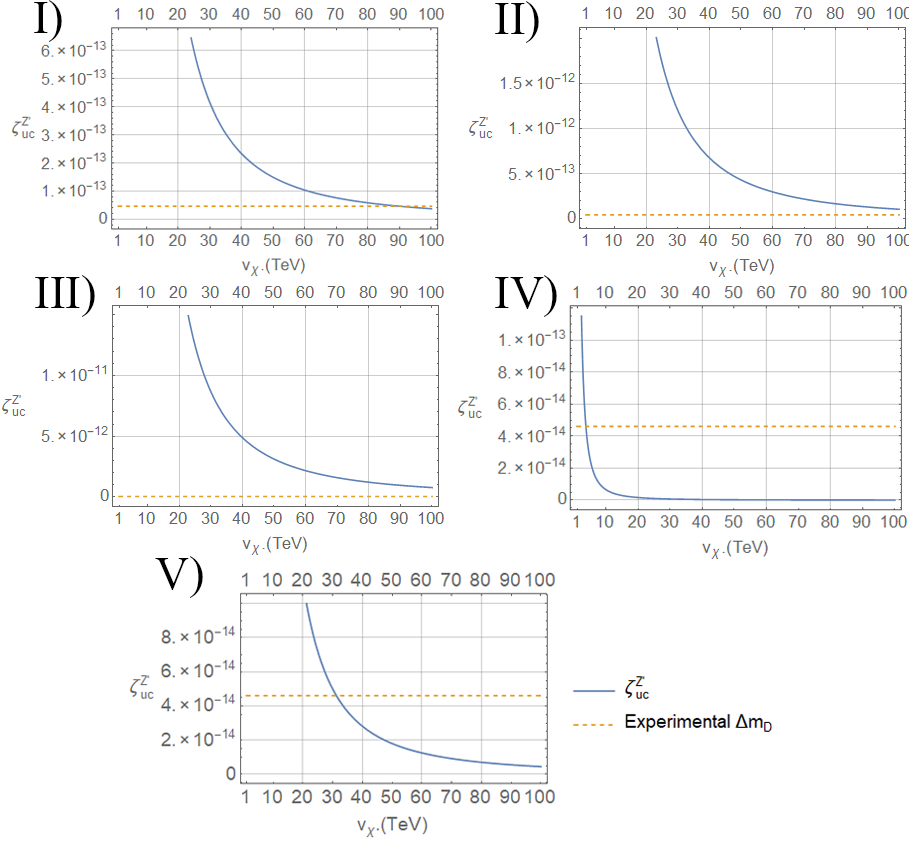}
	\caption{Contributions to $\Delta m_D$ from the $Z'$ boson. The numbers I-V correspond to the numerical solutions presented in appendix \ref{sec:numericalsolutionsused}.}
	\label{fig:ZetaZprime_uc_Todos}
\end{figure}

The $Z'$ contribution to the mass differences is mostly dependent on its mass, with a smaller dependence on the Yukawa constants and quark diagonalization matrices. It can be seen in Figs. \ref{fig:ZetaZprime_ds_Todos}-\ref{fig:ZetaZprime_uc_Todos} that the larger the $v_\chi$ the smaller this boson contribution. This is because $m_{Z'}$ is directly connect to $v_\chi$, the higher this vacuum expectation value, the higher the $Z'$ mass, and then, the smaller its contribution.

However, the Yukawa couplings and diagonalization matrices still play an important role. In most graphs in Figs. \ref{fig:ZetaZprime_ds_Todos}-\ref{fig:ZetaZprime_uc_Todos} the $Z'$ contribution is above the experimental value for all $v_\chi$ values, with the exceptions of Fig. \ref{fig:ZetaZprime_ds_Todos} I; Figs. \ref{fig:ZetaZprime_bd_Todos} IV and V; and Figs. \ref{fig:ZetaZprime_uc_Todos} I, IV and V. This is solely due to the Yukawa couplings and diagonalization matrices, which may increase or decrease this boson contribution depending on its values and signs.

\subsubsection{$h_1$ and $h_2$ scalars}

\begin{figure}
	\centering
	\includegraphics[width=\linewidth]{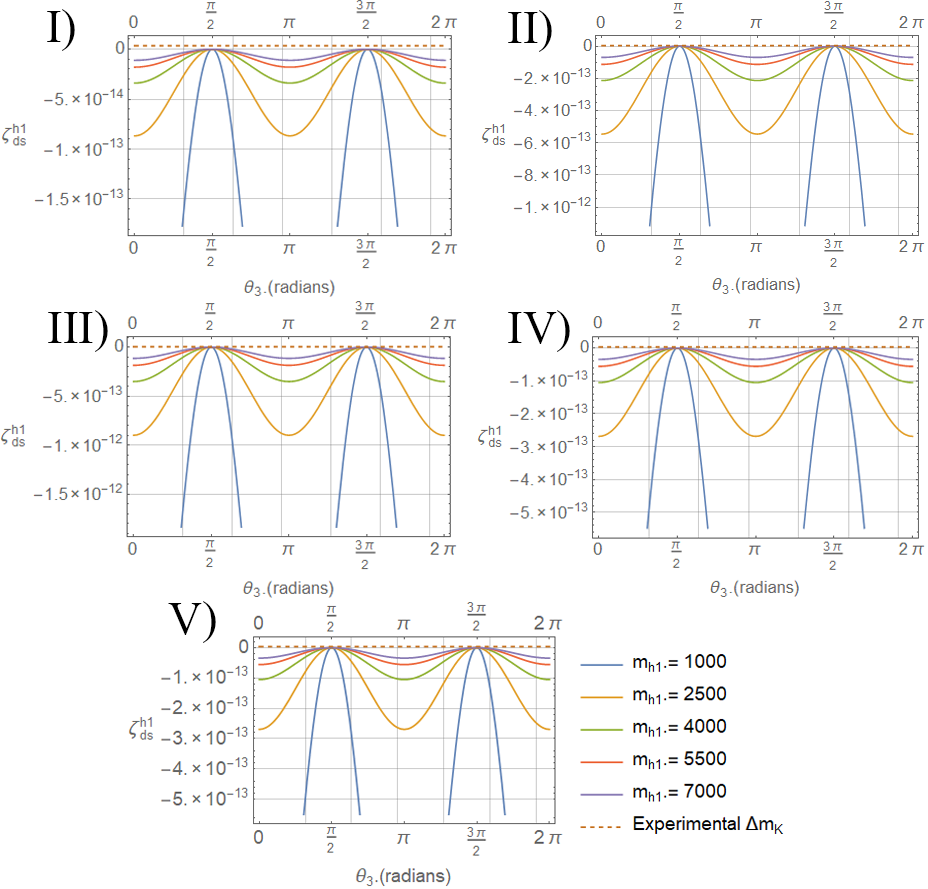}
	\caption{Contributions to $\Delta m_K$ from the $h_1$ scalar. The numbers I-V correspond to the numerical solutions presented in appendix \ref{sec:numericalsolutionsused}.}
	\label{fig:Zetah1_ds_Todos}
\end{figure}

\begin{figure}
	\centering
	\includegraphics[width=\linewidth]{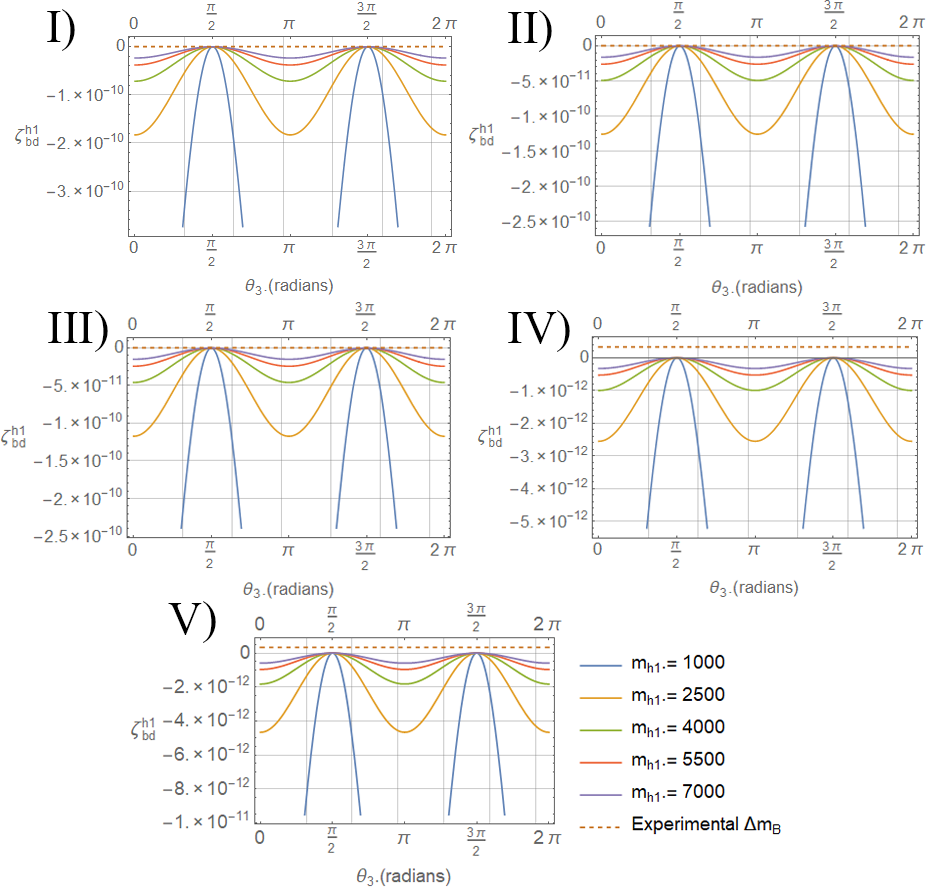}
	\caption{Contributions to $\Delta m_B$ from the $h_1$ scalar. The numbers I-V correspond to the numerical solutions presented in appendix \ref{sec:numericalsolutionsused}.}
	\label{fig:Zetah1_bd_Todos}
\end{figure}

\begin{figure}
	\centering
	\includegraphics[width=\linewidth]{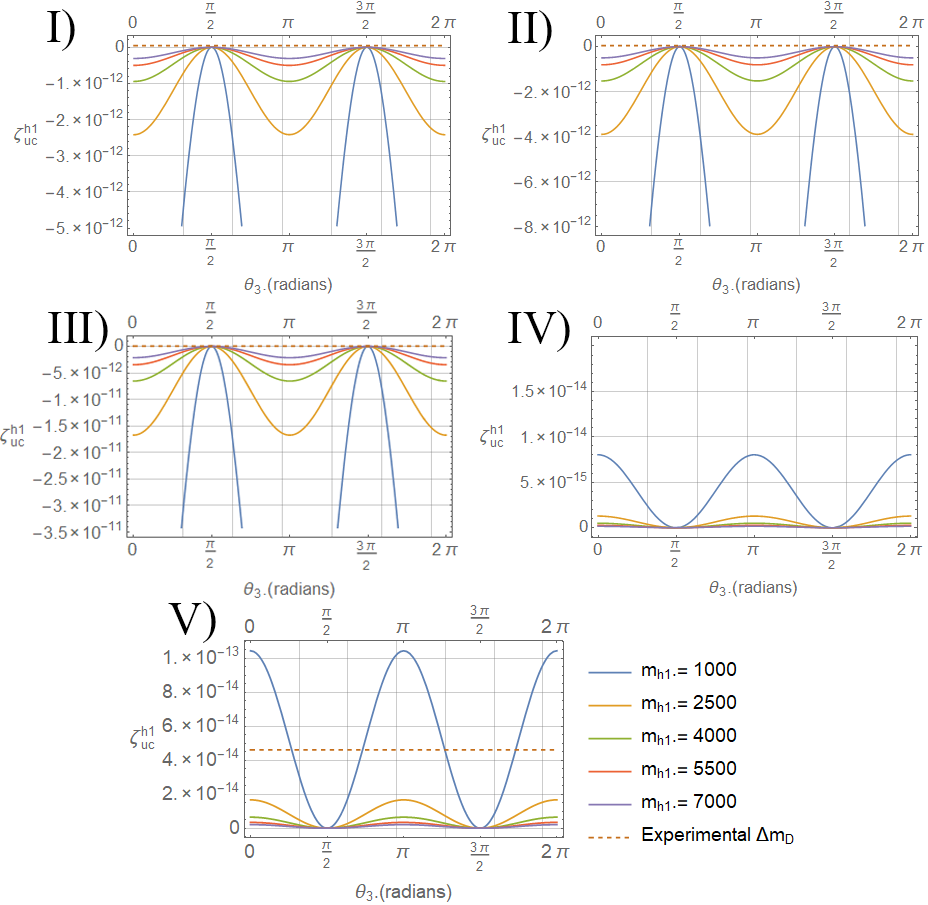}
	\caption{Contributions to $\Delta m_D$ from the $h_1$ scalar. The numbers I-V correspond to the numerical solutions presented in appendix \ref{sec:numericalsolutionsused}.}
	\label{fig:Zetah1_uc_Todos}
\end{figure}

\begin{figure}
	\centering
	\includegraphics[width=\linewidth]{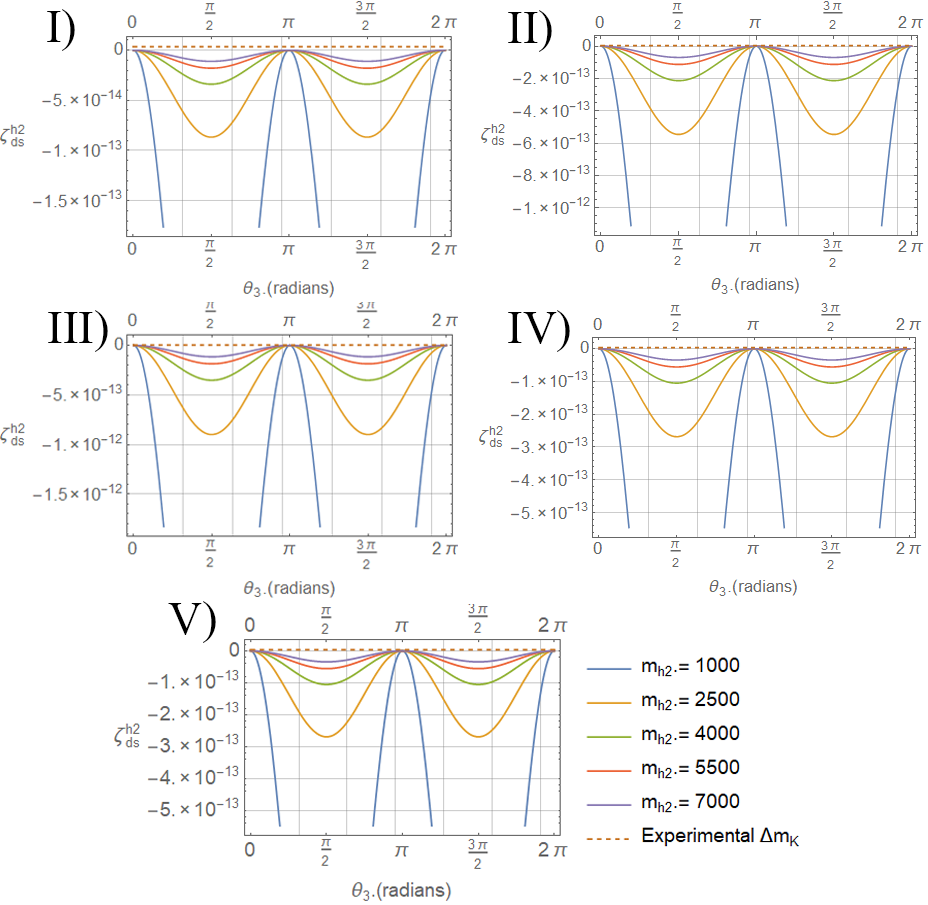}
	\caption{Contributions to $\Delta m_K$ from the $h_2$ scalar. The numbers I-V correspond to the numerical solutions presented in appendix \ref{sec:numericalsolutionsused}.}
	\label{fig:Zetah2_ds_Todos}
\end{figure}

\begin{figure}
	\centering
	\includegraphics[width=\linewidth]{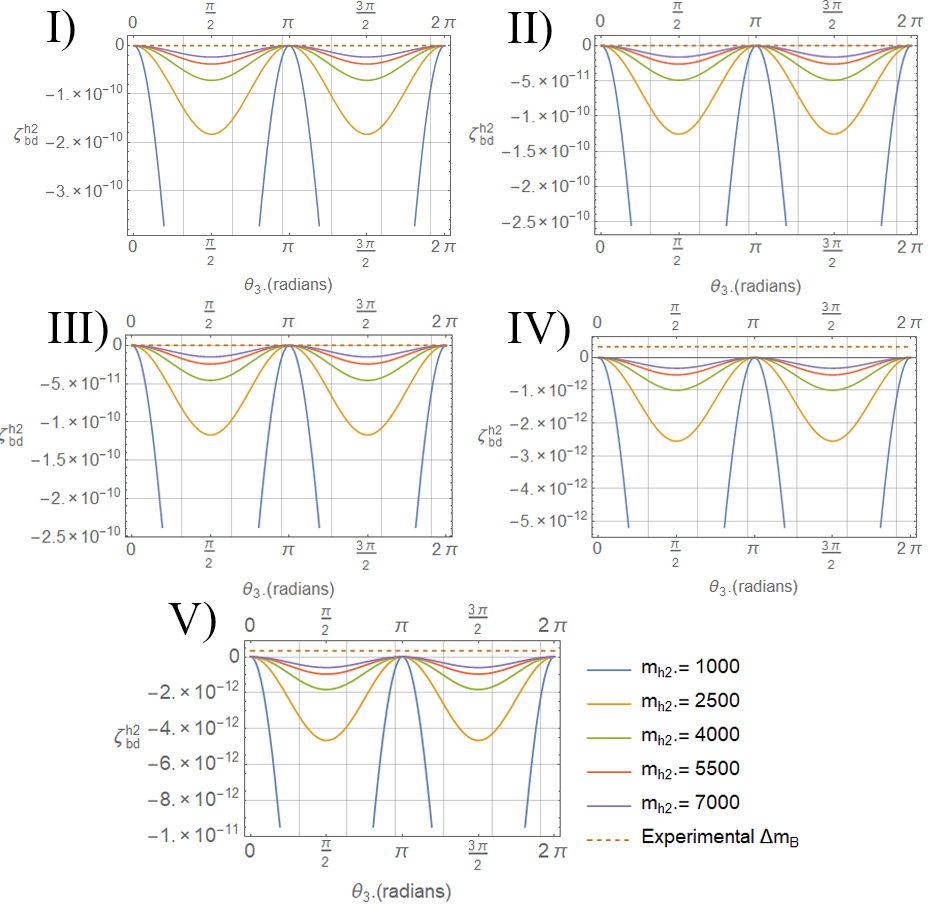}
	\caption{Contributions to $\Delta m_B$ from the $h_2$ scalar. The numbers I-V correspond to the numerical solutions presented in appendix \ref{sec:numericalsolutionsused}.}
	\label{fig:Zetah2_bd_Todos}
\end{figure}

\begin{figure}
	\centering
	\includegraphics[width=\linewidth]{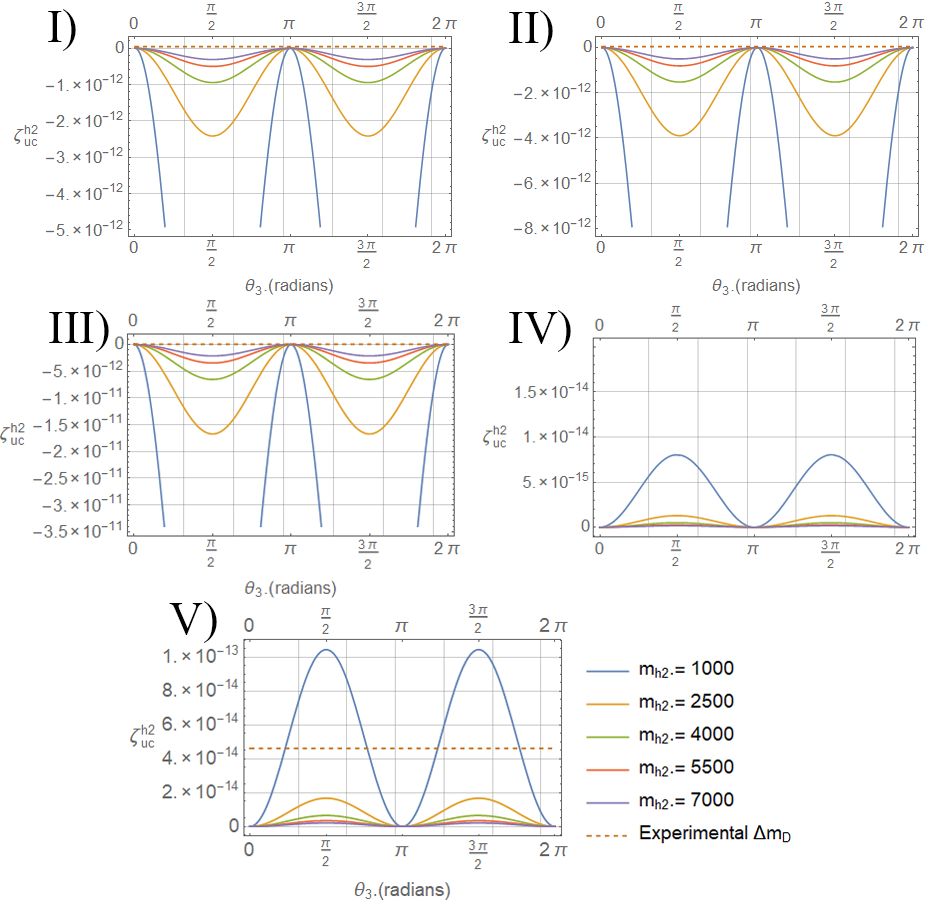}
	\caption{Contributions to $\Delta m_D$ from the $h_2$ scalar. The numbers I-V correspond to the numerical solutions presented in appendix \ref{sec:numericalsolutionsused}.}
	\label{fig:Zetah2_uc_Todos}
\end{figure}

These neutral CP-even scalars (Figs. \ref{fig:Zetah1_ds_Todos}-\ref{fig:Zetah2_uc_Todos}) have a similar behavior when we consider its masses, the higher the mass the lower its contribution (in magnitude). Whether a contribution is going to be positive or negative depends on the values and signs of the Yukawa couplings and diagonalization matrices, and it varies from solution to solution.

The masses of these scalars depend on all three vacuum expectation values ($v_\eta$, $v_\rho$ and $v_\chi$), but also depend on several unknown constants of the scalar potential, therefore leaving its masses completely free. That is why we chose to plot our graphs as a function of $\theta_3$, that controls the projection of the symmetry eigenstates over the mass eigenstates (see appendix \ref{sec:scalars2}). From the plots we can see that the $h_1$ contributions are maximal for $\theta_3=0,\pi,2\pi$, when $cos\theta_3=\pm1$; and the $h_2$ contributions are maximal for $\theta_3=\pi/2,3\pi/2$, when $sin\theta_3=\pm1$. This is obvious from Eq. \ref{eq:rotacaoescalaresCPpar}, where we can see that the projections of the $X^0_\eta$ and $X^0_\rho$ symmetry eigenstates over $h_1$ are proportional to $cos\theta_3$ and their projections over $h_2$ are proportional to $sin\theta_3$.

\subsubsection{$A^0$ scalar}

\begin{figure}
	\centering
	\includegraphics[width=\linewidth]{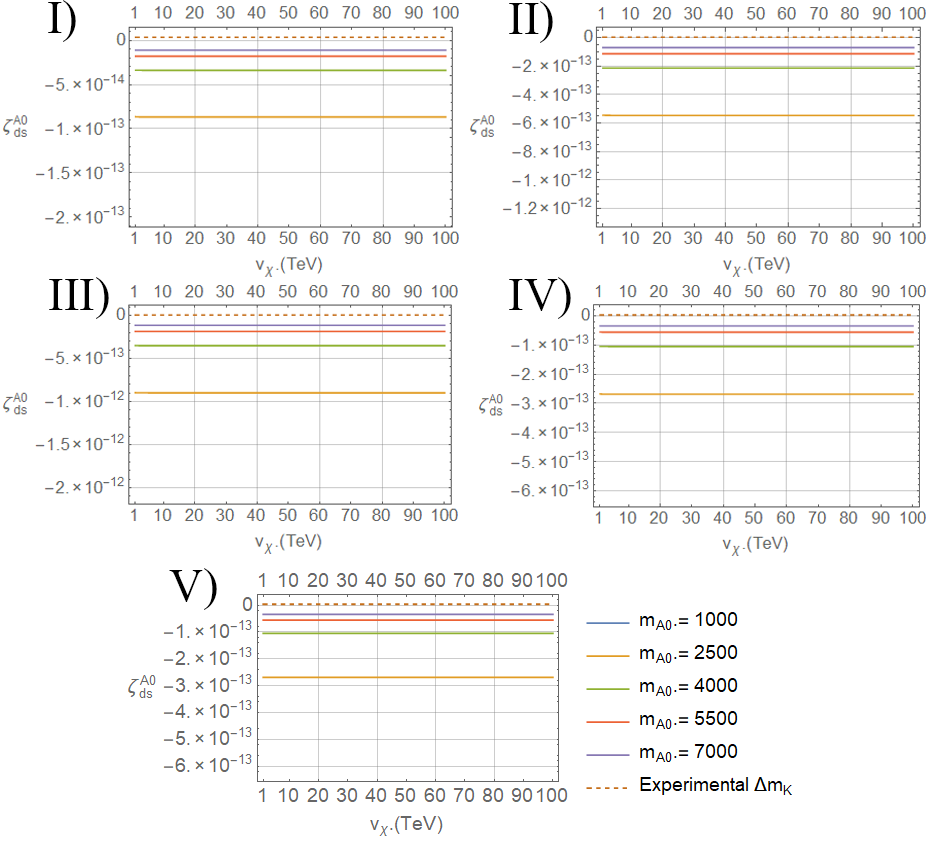}
	\caption{Contributions to $\Delta m_K$ from the $A^0$ scalar. The numbers I-V correspond to the numerical solutions presented in appendix \ref{sec:numericalsolutionsused}.}
	\label{fig:ZetaA0_ds_Todos}
\end{figure}

\begin{figure}
	\centering
	\includegraphics[width=\linewidth]{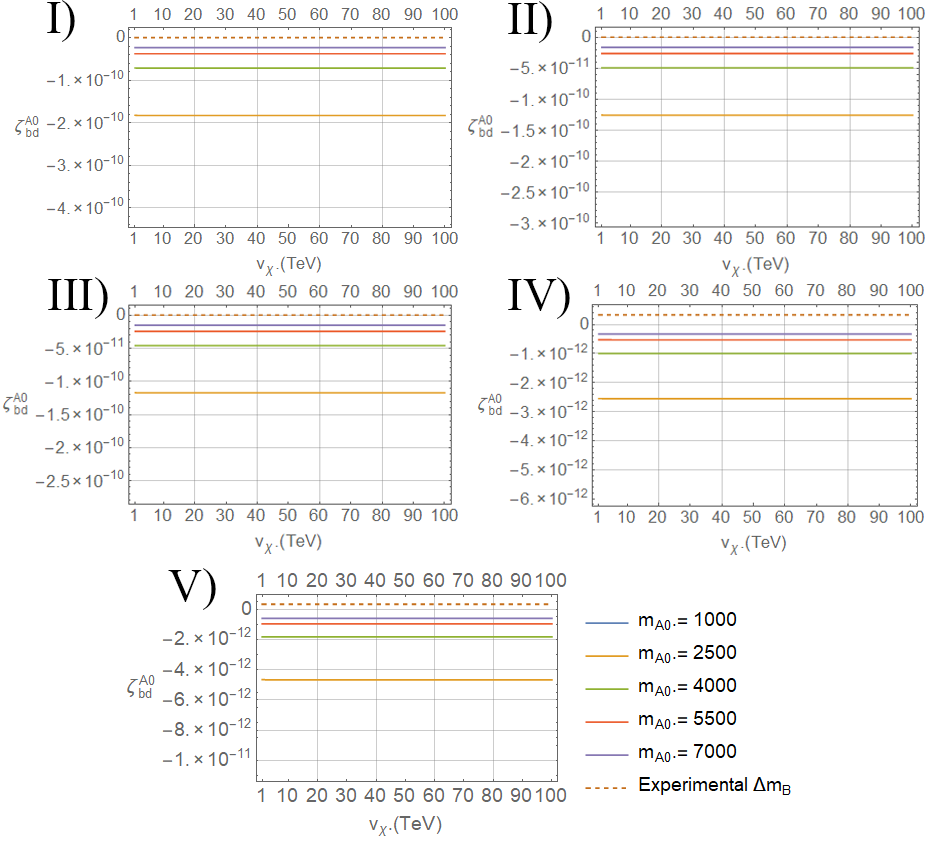}
	\caption{Contributions to $\Delta m_B$ from the $A^0$ scalar. The numbers I-V correspond to the numerical solutions presented in appendix \ref{sec:numericalsolutionsused}.}
	\label{fig:ZetaA0_bd_Todos}
\end{figure}

\begin{figure}
	\centering
	\includegraphics[width=\linewidth]{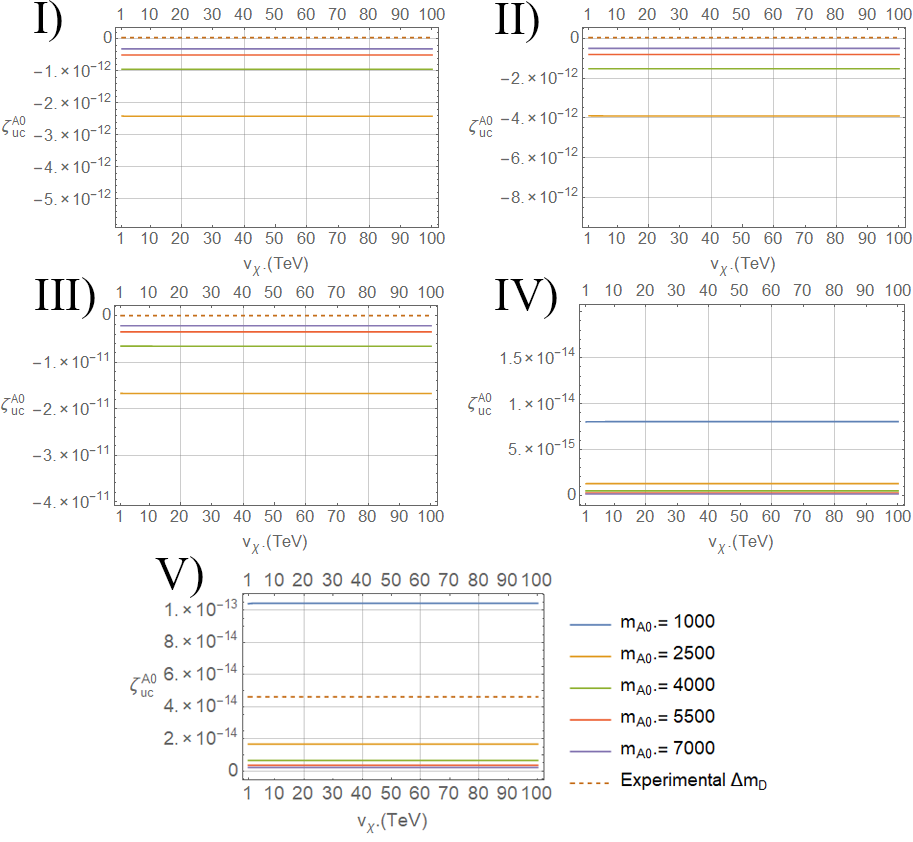}
	\caption{Contributions to $\Delta m_D$ from the $A^0$ scalar. The numbers I-V correspond to the numerical solutions presented in appendix \ref{sec:numericalsolutionsused}.}
	\label{fig:ZetaA0_uc_Todos}
\end{figure}

The contribution from this CP-odd scalar are shown in Figs. \ref{fig:ZetaA0_ds_Todos}-\ref{fig:ZetaA0_uc_Todos}. Similar to the previous case, the $A^0$ mass also depends on a constant from the scalar potential, which allows us to consider its mass to be a free parameter. However, the projections of $X^0_\eta$ and $X^0_\rho$ over $A^0$ depend on $v_\chi$ (see Eq. \ref{a4}), therefore we chose to plot varying this parameter. Notwithstanding, given that we are exploring values of $v_\chi$ much higher than the values of the other vacuum expectation values, the variation of the $A^0$ contribution with respect to this parameter is negligible.

Similar to the previous cases, higher masses imply lower contribution to the mass differences, but only in magnitude. Whether this contribution is going to be positive or negative once again depends on the values and signs of the Yukawa couplings and diagonalization matrices. With our numerical solutions we find that all contributions are negative, except for the ones in Figs. \ref{fig:ZetaA0_uc_Todos} IV and V.

\subsection{Numerical results for $\Delta m_K$, $\Delta m_B$ and $\Delta m_D$}\label{sec:numericalresults}

To find accordance between theoretical predictions and experimental results for the mass differences we have to satisfy, simultaneously, the following equations
\begin{eqnarray}
\Delta M_K^{331}+\Delta M_K^{SM}&=&\Delta M_K^{exp}, \nonumber \\
\Delta M_B^{331}+\Delta M_B^{SM}&=&\Delta M_B^{exp}, \\
\Delta M_D^{331}+\Delta M_D^{SM}&=&\Delta M_D^{exp}, \nonumber
\end{eqnarray}
where $\Delta M_X^{331}$ is given in sec. \ref{sec:theoreticalcalculations}. The experimental values for the mass differences are: $\Delta M_K^{exp}=(3.484 \pm 0.006) \times 10^{-15}$, $\Delta M_B^{exp}=(1.1691 \pm 0.0014)\times 10^{-11}$ and $\Delta M_D^{exp}=(6.52 \pm 2.90)\times 10^{-15}$, all in GeV \cite{PDG}. Taking into account the values and uncertainties for the SM predictions and the experimental values we end up with
\begin{eqnarray}
\Delta M_K^{331}&=&\Delta M_K^{exp}-\Delta M_K^{SM}= (3.201 \pm 0.013)\times 10^{-15}\, \textrm{GeV}\nonumber, \\
\Delta M_B^{331}&=&\Delta M_B^{exp}-\Delta M_B^{SM}=  (1.10 \pm 0.18)\times 10^{-11}\, \textrm{GeV}\label{eq:331maisSMigualEXP}, \\
\Delta M_D^{331}&=&\Delta M_D^{exp}-\Delta M_D^{SM}=  (6.52 \pm 2.90)\times 10^{-15}\, \textrm{GeV}\nonumber.
\end{eqnarray}
We must note that even the SM is not capable of explaining these mass differences. If it were able to do so, the differences in Eq. \ref{eq:331maisSMigualEXP} should be null within the uncertainties, noting that the $K^0-\bar{K}^0$ has the largest discrepancy.

The $\zeta^{331}_{ij}$ functions - that contribute to the $\Delta M_X^{331}$ - depend on 32 parameters, as discussed on sec. \ref{sec:individualcontributions}, which we have to explore. Of those, the 27 phases and angles from the diagonalization matrices are restricted to $[0,\,2\pi]$, and so is $\theta_3$. Meanwhile, the vacuum expectation value $v_\chi$ and the scalar masses are free.

To search for numerical solutions we did the following: fixed the scalar masses (values shown in Table \ref{table:massas}), generated random solutions for the diagonalization matrices (following the procedure described in appendix \ref{sec:numericalsolutionsquarkmassmatrices}) and plotted a graph varying $\theta_3$ from $0$ to $2 \pi$ and $v_\chi$ from $1$ TeV to $100$ TeV, which should show only the values for $v_\chi$ and $\theta_3$ where all equations in Eq. \ref{eq:331maisSMigualEXP} were satisfied simultaneously within three standard deviations (3$\sigma$).

\begin{table}[ht]
	\centering
	\begin{tabular}{|c|c|c|}\hline
		$m_{h1}$ & $m_{h2}$ & $m_{A0}$ \\ \hline
		250 & 500 & 750 \\ \hline
		750 & 500 & 250 \\ \hline
		500 & 750 & 250 \\ \hline
		800 & 900 & 700 \\ \hline
		1000 & 1000 & 1000 \\ \hline
		2500 & 4000 & 5000 \\ \hline
		3000 & 3000 & 3000 \\ \hline
		12000 & 20000 & 15000 \\ \hline
		25000 & 40000 & 55000 \\ \hline
		30000 & 30000 & 30000 \\ \hline
		40000 & 20000 & 15000 \\ \hline
		50000 & 90000 & 70000 \\ \hline
		70000 & 60000 & 50000 \\ \hline
		125000 & 140000 & 155000 \\ \hline
	\end{tabular}
\caption{Set of scalar masses values explored. Each row corresponds to a set of values used to search for solutions of Eq. \ref{eq:331maisSMigualEXP}.}
\label{table:massas}
\end{table}
For each set of values presented in Table \ref{table:massas} we generated one thousand different numerical solutions for the diagonalization matrices, plotting in \textsf{Mathematica 10} using the command \textsf{RegionPlot} with the setting \textsf{PlotPoints$\rightarrow$200}. Our program checked each graph for plotted points, not finding anything in all our attempts (corresponding to 14000 attempts in total, 1000 for each set of masses). Therefore, we are led to believe that when all contributions are taken into account simultaneously, there are no solutions for Eq. \ref{eq:331maisSMigualEXP} in the $1$ TeV$<v_\chi<100$ TeV range. If there are solutions in this range, they must be in very small regions of the $\theta_3 \times v_\chi$ parameter space, or they are possible only for other sets of values for the scalar masses.

\newpage
\section{Conclusions}

In this work we are able to show analytical solutions for the mass eigenstates of the CP-even neutral sector and it's masses of the m331 scalar potential. So, we now have the scalar sector of the model completely solved. This was done by imposing that the Higgs-fermion interactions in the m331 are the same as the ones in the SM at tree level. With that we have also shown that the Higgs self-interactions in the m331 are the same as the SM, reinforcing the idea that the m331 can properly reproduce and extend the SM. We also presented a method to find numerical solutions for the fermion mass and diagonalization matrices, something never shown before in the literature, now allowing us to make complete analyses using the m331 without leaving these parameters free.

Finally, we were able to show analytical solutions for all m331 contributions to the mass differences in the $K^0-\bar{K}^0$, $B_s^0-\bar{B}_s^0$ and $D^0-\bar{D}^0$ systems, at tree level. As for numerical solutions for the mass differences, we are led to believe that the m331 is unable to provide those in the parameter range explored, or if there is a solution, it is in a very small region of the parameter space. Therefore, the possibility of a solution at higher values of $v_\chi$ remains. We must remark that the SM is also unable to explain these mass differences (see Eq. \ref{eq:331maisSMigualEXP}), which suggests the existence of new physical effects in action.

Previous works have presented solutions for the $K^0-\bar{K}^0$, $B_s^0-\bar{B}_s^0$ and $D^0-\bar{D}^0$ systems \cite{Machado:2016jzb,Queiroz:2016gif}. However, these works have not considered all contributions that the m331 has at tree level, the first considered the $Z'$ boson and the $A^0$ CP-odd scalar (studied individually), while the second considered only the $Z'$ boson. When considering only a single contribution the analysis becomes simpler, and we even found solutions when considering only the $Z'$ boson (Figs. \ref{fig:ZetaZprime_ds_Todos}-\ref{fig:ZetaZprime_uc_Todos}). However, when considering all contributions - $Z'$, $A^0$, $h_1^0$ and $h_2^0$ - and imposing that the sum of these contributions at a given set of parameter values must satisfy all three mass differences simultaneously, the problem becomes much harder. Therefore, the analysis we did is more difficult than the ones we mentioned, it is more complete and provides a harsher test for the model studied. 

\acknowledgments

G. De Conto would like to thank Conselho Nacional de Desenvolvimento Cient\'{i}fico e Tecnol\'{o}gico (CNPq) for financial support, A. C. B. Machado would like to thank Coordena\c{c}\~{a}o de Aperfei\c{c}oamento de Pessoal de N\'{i}vel Superior (CAPES) for financial support and J. P. B. C. de Melo would like to thank Funda\c{c}\~{a}o de Amparo a Pesquisa do Estado de S\~{a}o Paulo (FAPESP), grant no. 2015/16295-5, and also CNPq.

\newpage
\appendix

\section{Neutral CP-odd scalars}\label{sec:CPodd}

The relation between the mass and symmetry eigenstates of the CP-odd scalars is given by

\begin{equation}
\left(\begin{array}{c}
I^0_\eta \\ I^0_\rho \\ I^0_\chi
\end{array}\right)=
\left(\begin{array}{ccc}
\frac{N_a}{|v_\chi|} & - \frac{N_b|v_\eta||v_\chi|}{|v_\rho|(|v_\eta|^2+|v_\chi|^2)} & \frac{N_c}{|v_\eta|} \\
0 & \frac{N_b}{|v_\chi|} & \frac{N_c}{|v_\rho|} \\
-\frac{N_a}{|v_\eta|} & - \frac{N_b|v_\eta|^2}{|v_\rho|(|v_\eta|^2+|v_\chi|^2)} & \frac{N_c}{|v_\chi|}
\end{array}\right)
\left(\begin{array}{c}
G^0_1 \\ G^0_2 \\ A^0
\end{array}\right)
\label{a4}
\end{equation}
where $G^0_1$ and $^0_2$ are the Goldstone bosons that give mass to the $Z$ and $Z'$ gauge bosons and $A^0$ is a massive eigenstate. The masses of these scalars are
\begin{equation}
m^2_{G_1^0}=m^2_{G_2^0}=0,\quad
m^2_{A^0}=A\left(\frac{1}{|v_\chi|^2}+\frac{1}{|v_\rho|^2}+\frac{1}{|v_\eta|^2}\right),
\end{equation}
where
\begin{eqnarray}
&&N_a=\left(\frac{1}{|v_\chi|^2}+\frac{1}{|v_\eta|^2}\right)^{-1/2},
\quad
N_b=\left(
\frac{1}{|v_\chi|^2}+\frac{|v_\eta|^2}{|v_\rho|^2(|v_\eta|^2+|v_\chi|^2)}\right)^{-1/2},
\nonumber \\&&
N_c=\left(\frac{1}{|v_\chi|^2}+\frac{1}{|v_\rho|^2}+\frac{1}{|v_\eta|^2}\right)^{-1/2}.
\label{a5}
\end{eqnarray}

\section{Numerical solutions used for the quark mass matrices}\label{sec:numericalsolutionsused}

In this appendix we present the numerical solutions used for the diagonalization and mass matrices throughout our work. They have been found using the procedure described in appendix \ref{sec:numericalsolutionsquarkmassmatrices} and follow the same notation.

\subsection{1st set of numerical solutions}

\begin{equation}
V^U_L=\left(
\begin{array}{ccc}
-0.351453-0.733396 i & 0.180442\, +0.242749 i & -0.206303-0.452288 i \\
-0.207082-0.432127 i & -0.272599-0.759689 i & 0.260085\, +0.226496 i \\
0.186328\, +0.272537 i & -0.0285573-0.506221 i & -0.248724-0.756351 i \\
\end{array}
\right)
\end{equation}

\begin{equation}
V^U_R=
\left(
\begin{array}{ccc}
0.114008\, -0.0862582 i & -0.407525+0.308332 i & 0.842955\, +0.0885602 i \\
-0.124017+0.0938303 i & 0.804546\, -0.223297 i & 0.472219\, +0.235942 i \\
-0.341126+0.915975 i & -0.11502+0.168727 i & 0.0274397\, -0.0465952 i \\
\end{array}
\right)
\end{equation}

\begin{equation}
V^D_L=\left(
\begin{array}{ccc}
-0.29525-0.614518 i & 0.238769\, +0.403548 i & -0.259166-0.498888 i \\
-0.288733-0.597706 i & -0.223829-0.664759 i & 0.217269\, +0.150083 i \\
0.179848\, +0.252665 i & -0.0404451-0.53675 i & -0.236675-0.748188 i \\
\end{array}
\right)
\end{equation}

\begin{equation}
V^D_R=\left(
\begin{array}{ccc}
-0.237633+0.702967 i & 0.00145773\, +0.58581 i & 0.308112\, -0.106108 i \\
0.0233675\, -0.0691261 i & 0.0547041\, +0.550733 i & -0.747995+0.358997 i \\
0.640829\, +0.182708 i & -0.591557+0.0241872 i & 0.177781\, +0.416924 i \\
\end{array}
\right)
\end{equation}

\begin{equation}
G=
\left(
\begin{array}{ccc}
0.843586\, +1.1935 i & 0.109032\, +0.297939 i & -0.0411038-0.0680892 i \\
-2.06044-0.905998 i & -0.374213-0.263531 i & 0.0931705\, +0.0788665 i \\
0 & 0 & 0 \\
\end{array}
\right)
\end{equation}

\begin{equation}
F=\left(
\begin{array}{ccc}
0 & 0 & 0 \\
0 & 0 & 0 \\
-0.620589-0.495472 i & -0.0998621-0.133434 i & 0.0303255\, +0.0326737 i \\
\end{array}
\right)
\end{equation}

\begin{equation}
\tilde{G}=\left(
\begin{array}{ccc}
0.00398493\, -0.00316975 i & -0.00267327+0.00371303 i & 0.00338226\, +0.000438449 i \\
-0.00302469+0.00831496 i & 0.000064437\, -0.00788955 i & -0.00572811+0.00160747 i \\
0 & 0 & 0 \\
\end{array}
\right)
\end{equation}

\begin{equation}
\tilde{F}=\left(
\begin{array}{ccc}
0 & 0 & 0 \\
0 & 0 & 0 \\
-0.0316173+0.0476692 i & 0.013544\, -0.0488196 i & -0.0390278+0.00425517 i \\
\end{array}
\right)
\end{equation}

\subsection{2nd set of numerical solutions}

\begin{equation}
V^U_L=
\left(
\begin{array}{ccc}
-0.123312+0.319625 i & -0.251103+0.525058 i & -0.167749+0.718161 i \\
0.337651\, -0.875194 i & -0.128595+0.213006 i & -0.0361346+0.238356 i \\
0.0273901\, +0.0437157 i & -0.0928577+0.76859 i & -0.0742078-0.626482 i \\
\end{array}
\right)
\end{equation}

\begin{equation}
V^U_R=
\left(
\begin{array}{ccc}
-0.338892-0.93851 i & -0.0114054-0.030299 i & 0.0244975\, -0.0519992 i \\
0.0166585\, +0.0461332 i & 0.0529072\, -0.894529 i & -0.0194015-0.440723 i \\
-0.0235509+0.037306 i & -0.438581-0.0601503 i & 0.89498\, +0.0330521 i \\
\end{array}
\right)
\end{equation}

\begin{equation}
V^D_L=\left(
\begin{array}{ccc}
-0.1962+0.509201 i & -0.2191+0.469707 i & -0.15384+0.640817 i \\
0.300153\, -0.782741 i & -0.178082+0.294152 i & -0.0699545+0.420056 i \\
0.0401336\, +0.00747276 i & -0.100227+0.777136 i & -0.0783129-0.61639 i \\
\end{array}
\right)
\end{equation}

\begin{equation}
V^D_R=
\left(
\begin{array}{ccc}
0.127195\, -0.93902 i & -0.265349-0.15974 i & 0.0764837\, -0.0169141 i \\
0.0426325\, -0.314736 i & 0.796843\, +0.493234 i & -0.123916+0.0743615 i \\
0.0229497\, -0.025687 i & 0.0464891\, -0.15386 i & -0.673059-0.72109 i \\
\end{array}
\right)
\end{equation}

\begin{equation}
G=
\left(
\begin{array}{ccc}
0.00329587\, +0.0103259 i & -0.0396832+0.0709601 i & 0.130413\, -0.178908 i \\
0.140219\, +0.0661044 i & -0.0315635+1.55791 i & -0.264804-3.13223 i \\
0 & 0 & 0 \\
\end{array}
\right)
\end{equation}

\begin{equation}
F=
\left(
\begin{array}{ccc}
0 & 0 & 0 \\
0 & 0 & 0 \\
-0.0220007-0.0179214 i & 0.0700629\, -0.275733 i & -0.0897047+0.569918 i \\
\end{array}
\right)
\end{equation}

\begin{equation}
\tilde{G}=
\left(
\begin{array}{ccc}
0.000148801\, -0.00006044 i & -0.0000654252+0.000276086 i & -0.000852132-0.000631743 i \\
-0.000617683-0.000347394 i & -0.00305859-0.000685908 i & -0.0121173+0.0146756 i \\
0 & 0 & 0 \\
\end{array}
\right)
\end{equation}

\begin{equation}
\tilde{F}=
\left(
\begin{array}{ccc}
0 & 0 & 0 \\
0 & 0 & 0 \\
0.00112207\, +0.00178691 i & 0.0103525\, +0.00356181 i & 0.0545233\, -0.0391231 i \\
\end{array}
\right)
\end{equation}

\subsection{3rd set of numerical solutions}

\begin{equation}
V^U_L=\left(
\begin{array}{ccc}
0.743237\, +0.38557 i & -0.0803848-0.120118 i & 0.311942\, +0.425132 i \\
0.463734\, +0.240572 i & 0.303175\, +0.391412 i & -0.403085-0.565226 i \\
0.129625\, +0.0959617 i & -0.455804-0.72542 i & -0.250385-0.421076 i \\
\end{array}
\right)
\end{equation}

\begin{equation}
V^U_R=
\left(
\begin{array}{ccc}
0.40963\, -0.579876 i & 0.252002\, -0.0330819 i & -0.61903+0.219429 i \\
0.36931\, -0.522799 i & -0.614665-0.0326903 i & 0.416892\, -0.193933 i \\
0.282601\, -0.0798464 i & 0.472691\, +0.577133 i & 0.524791\, +0.286072 i \\
\end{array}
\right)
\end{equation}

\begin{equation}
V^D_L=
\left(
\begin{array}{ccc}
0.620589\, +0.322785 i & -0.147959-0.212767 i & 0.394231\, +0.537423 i \\
0.614101\, +0.317391 i & 0.296204\, +0.384393 i & -0.312194-0.43765 i \\
0.148328\, +0.108847 i & -0.442954-0.709643 i & -0.268136-0.442917 i \\
\end{array}
\right)
\end{equation}

\begin{equation}
V^D_R=
\left(
\begin{array}{ccc}
0.19149\, +0.0780316 i & 0.24068\, +0.940359 i & 0.105803\, +0.0620162 i \\
0.90497\, +0.368773 i & -0.0566437-0.200328 i & 0.0204335\, -0.0357594 i \\
-0.046864-0.00907322 i & -0.120163-0.00433824 i & 0.769787\, -0.625053 i \\
\end{array}
\right)
\end{equation}

\begin{equation}
G=
\left(
\begin{array}{ccc}
0.132938\, -0.181065 i & 0.519405\, +0.13802 i & 0.437985\, -0.0665578 i \\
-0.324656+1.08492 i & -2.88315+0.369877 i & -2.02475+1.128 i \\
0 & 0 & 0 \\
\end{array}
\right)
\end{equation}

\begin{equation}
F=
\left(
\begin{array}{ccc}
0 & 0 & 0 \\
0 & 0 & 0 \\
-0.0368042+0.145005 i & -0.366832+0.053147 i & -0.257499+0.154795 i \\
\end{array}
\right)
\end{equation}

\begin{equation}
\tilde{G}=
\left(
\begin{array}{ccc}
0.000185154\, +0.0000580992 i & -0.000492897+0.000261814 i & 0.00113971\, -0.00436321 i \\
0.000898032\, -0.000852909 i & 0.00132768\, -0.00207653 i & 0.00252163\, +0.0202648 i \\
0 & 0 & 0 \\
\end{array}
\right)
\end{equation}

\begin{equation}
\tilde{F}=
\left(
\begin{array}{ccc}
0 & 0 & 0 \\
0 & 0 & 0 \\
0.000725828\, -0.00131606 i & 0.00407507\, -0.00557467 i & 0.00774354\, +0.0557173 i \\
\end{array}
\right)
\end{equation}

\subsection{4th set of numerical solutions}

\begin{equation}
V^U_L=
\left(
\begin{array}{ccc}
-0.0401711-0.998995 i & 0.0126172\, +0.0148803 i & 0.00368908\, -0.000626611 i \\
-0.000210393-0.00523216 i & -0.0745351-0.353602 i & 0.404742\, +0.83998 i \\
0.00125312\, -0.0191215 i & -0.562711-0.743227 i & -0.280635-0.2277 i \\
\end{array}
\right)
\end{equation}

\begin{equation}
V^U_R=
\left(
\begin{array}{ccc}
-0.261258-0.0373503 i & -0.0127772+0.0826928 i & -0.933499+0.227877 i \\
-0.918085-0.131252 i & -0.189101-0.216412 i & 0.236342\, -0.0379802 i \\
0.254522\, +0.0739013 i & -0.57321-0.762781 i & -0.114698+0.0787177 i \\
\end{array}
\right)
\end{equation}

\begin{equation}
V^D_L=
\left(
\begin{array}{ccc}
-0.0390287-0.972598 i & 0.026989\, +0.0862009 i & -0.0891539-0.192731 i \\
-0.00931023-0.229452 i & -0.0465338-0.31054 i & 0.406897\, +0.827921 i \\
0.00457062\, -0.0205907 i & -0.565828-0.757787 i & -0.263895-0.19296 i \\
\end{array}
\right)
\end{equation}

\begin{equation}
V^D_R=
\left(
\begin{array}{ccc}
-0.432283-0.209645 i & -0.0486966+0.184844 i & -0.512853-0.685291 i \\
0.436318\, +0.211601 i & 0.53197\, -0.546014 i & -0.18476-0.386772 i \\
0.102349\, +0.723572 i & -0.616354-0.0494584 i & 0.0411379\, -0.286246 i \\
\end{array}
\right)
\end{equation}

\begin{equation}
G=
\left(
\begin{array}{ccc}
-0.00493106+0.022322 i & 0.0629388\, -0.0540882 i & -0.00748637-0.00951652 i \\
-0.895006+0.658952 i & 4.03786\, +0.0128217 i & 0.0272439\, -0.584745 i \\
0 & 0 & 0 \\
\end{array}
\right)
\end{equation}

\begin{equation}
F=
\left(
\begin{array}{ccc}
0 & 0 & 0 \\
0 & 0 & 0 \\
-0.0936976+0.0433784 i & 0.339516\, +0.0858041 i & 0.0150379\, -0.0508102 i \\
\end{array}
\right)
\end{equation}

\begin{equation}
\tilde{G}=
\left(
\begin{array}{ccc}
-0.000378658+0.000176646 i & 0.000018025\, -0.000248531 i & 0.000219866\, -0.000046429 i \\
-0.0149809-0.00810275 i & 0.00959465\, -0.010708 i & 0.00483943\, +0.00473589 i \\
0 & 0 & 0 \\
\end{array}
\right)
\end{equation}

\begin{equation}
\tilde{F}=
\left(
\begin{array}{ccc}
0 & 0 & 0 \\
0 & 0 & 0 \\
-0.0173535-0.0194338 i & 0.0182604\, -0.0132425 i & 0.00389669\, +0.00912257 i \\
\end{array}
\right)
\end{equation}

\subsection{5th set of numerical solutions}

\begin{equation}
V^U_L=
\left(
\begin{array}{ccc}
-0.355121-0.899742 i & 0.069077\, +0.243693 i & 0.00895344\, -0.0107722 i \\
0.0167247\, -0.106206 i & 0.0617245\, -0.368669 i & -0.857235-0.337434 i \\
-0.0647165-0.220469 i & -0.14936-0.879663 i & 0.388485\, +0.013026 i \\
\end{array}
\right)
\end{equation}

\begin{equation}
V^U_R=
\left(
\begin{array}{ccc}
-0.0838139-0.600685 i & -0.0704808-0.765623 i & 0.0757563\, +0.187798 i \\
0.768063\, +0.152261 i & -0.525125-0.164984 i & 0.252616\, -0.141793 i \\
0.0768798\, -0.114624 i & -0.000760595+0.325389 i & 0.492631\, +0.795228 i \\
\end{array}
\right)
\end{equation}

\begin{equation}
V^D_L=
\left(
\begin{array}{ccc}
-0.349658-0.854982 i & 0.0551556\, +0.312875 i & 0.205062\, +0.0669666 i \\
-0.061058-0.297169 i & 0.0819\, -0.268017 i & -0.849509-0.331831 i \\
-0.061386-0.227065 i & -0.147553-0.894399 i & 0.35309\, -0.000906843 i \\
\end{array}
\right)
\end{equation}

\begin{equation}
V^D_R=
\left(
\begin{array}{ccc}
0.479254\, -0.876087 i & -0.0105637-0.0102439 i & 0.0228135\, +0.0452909 i \\
-0.0067006+0.00964172 i & -0.918423+0.2755 i & -0.196558+0.204516 i \\
0.0484323\, +0.0174558 i & -0.279603+0.0469585 i & 0.743228\, -0.603804 i \\
\end{array}
\right)
\end{equation}

\begin{equation}
G=
\left(
\begin{array}{ccc}
0.0919884\, +0.113354 i & -0.324862-0.0982262 i & -0.939293+0.260041 i \\
0.404999\, +0.394205 i & -1.29696-0.230302 i & -3.50474+1.42979 i \\
0 & 0 & 0 \\
\end{array}
\right)
\end{equation}

\begin{equation}
F=
\left(
\begin{array}{ccc}
0 & 0 & 0 \\
0 & 0 & 0 \\
0.0236269\, -0.0455047 i & 0.00782483\, +0.12876 i & 0.204637\, +0.310318 i \\
\end{array}
\right)
\end{equation}

\begin{equation}
\tilde{G}=
\left(
\begin{array}{ccc}
-0.000155785+0.000263289 i & 0.000146042\, -0.00179712 i & 0.00222458\, +0.00503009 i \\
-0.000569328+0.000997344 i & -0.000101858-0.00645537 i & 0.0105613\, +0.0185474 i \\
0 & 0 & 0 \\
\end{array}
\right)
\end{equation}

\begin{equation}
\tilde{F}=
\left(
\begin{array}{ccc}
0 & 0 & 0 \\
0 & 0 & 0 \\
0.0018815\, +0.000626997 i & -0.00909882+0.000446632 i & 0.0290354\, -0.0238586 i \\
\end{array}
\right)
\end{equation}

\newpage

\newpage




\end{document}